\documentclass[10pt,aps,pra,twocolumn,superscriptaddress,floatfix,nofootinbib]{revtex4-2}

\makeatletter
\def\@bibdataout@aps{%
 \immediate\write\@bibdataout{%
  @CONTROL{%
   apsrev41Control,author="08",editor="1",pages="0",title="0",year="1",eprint="1"%
  }%
 }%
 \if@filesw
  \immediate\write\@auxout{\string\citation{apsrev41Control}}%
 \fi
}%
\makeatother % Phew.

\usepackage[T1]{fontenc}
\usepackage{amsfonts}
\usepackage{braket}
\usepackage{mathrsfs}
\usepackage{mathtools}
\usepackage[breaklinks=true]{hyperref}
\usepackage[usenames,dvipsnames]{color}
\hypersetup{
  colorlinks   = true, %Colours links instead of ugly boxes
  urlcolor     = blue, %Colour for external hyperlinks
  linkcolor    = blue, %Colour of internal links
  citecolor   = red %Colour of citations
}
\usepackage{cleveref}

\usepackage{tikz}
\usetikzlibrary{quantikz}

\newcommand{\half}{\tfrac{1}{2}}
\newcommand{\smallfrac}[2]{\mbox{$\frac{#1}{#2}$}}
\newcommand{\N}{\mathscr{N}}
\newcommand{\betalim}{|\beta| \rightarrow \infty}

\newcommand{\dg}{^\dagger}
\newcommand\Tr{\mathrm{Tr}}
\newcommand{\ip}[2]{\langle{#1}|{#2}\rangle} % inner product
\newcommand{\op}[2]{\ket{#1}\!\bra{#2}}        % outer product
\newcommand{\expt}[1]{\langle{#1}\rangle}    % expectation

\newcommand{\focksym}{N}    % expectation

\newcommand\Id{\mathbb{I}}

\newcommand{\sch}{Schr\"odinger}

\DeclareMathOperator{\sech}{sech}

\crefformat{equation}{Eq.~(#2#1#3)} % These change 'equation' to Eq., more PRA-style
\crefformat{section}{Sec.~#2#1#3} % These change 'equation' to Eq., more PRA-style
\Crefformat{equation}{Equation~(#2#1#3)}
\crefformat{figure}{Fig.~#2#1#3}
\crefrangeformat{equation}{Eqs.~#3(#1)#4--#5(#2)#6}
\crefmultiformat{equation}{Eqs.~(#2#1#3)}{ and~(#2#1#3)}{, (#2#1#3)}{ and~(#2#1#3)}

\makeatother

\begin{document}

% Limit for box width tolerances
\hfuzz=150pt
\hbadness=10000

\title{
Homodyne measurement with a \sch\ cat state as a local oscillator
%Homodyne measurement with superposition of coherent states as a local oscillator
}

\author{Joshua Combes}
\email[]{joshua.combes@gmail.com}
\address{Department of Electrical, Computer, and Energy Engineering, University of Colorado Boulder, Colorado 80309, USA}
%\affiliation{School of Mathematics \& Physics, The University of Queensland, St Lucia QLD, Australia}

\author{Austin P. Lund}
\email[]{a.lund@uq.edu.au}

\affiliation{Dahlem Center for Complex Quantum Systems, Freie Universit\"at Berlin, 14195 Berlin, Germany}
\affiliation{Centre for Quantum Computation and Communications Technology, School of Mathematics and Physics, The University of Queensland, St Lucia QLD, Australia}

\begin{abstract}
Homodyne measurements are a widely used quantum measurement. Using a coherent state of large amplitude as the local oscillator, it can be shown that the quantum homodyne measurement limits to a field quadrature measurement. In this work, we give an example of a general idea: injecting non-classical states as a local oscillator can led to non-classical measurements. Specifically we consider injecting a superposition of coherent states, a Schrödinger cat state, as a local oscillator. We derive the Kraus operators and the positive operator-valued measure (POVM) in this situation.  
\end{abstract}
\maketitle

\section{Introduction}
Homodyne measurement is a low noise, high sensitivity technique to detect a quadrature of the electromagnetic field.  This is achieved by the mixing of a high power, phase stable local oscillator with an input signal and detecting the resulting low frequency components.  
In practice, homodyne detection can in operate very close to the noise limits imposed by quantum mechanics~\cite{HuntingtonLovovsky2012}.  Hence homodyne measurements have become a vital component of optical and microwave quantum-- optics, communication, and computation. 

\begin{figure}[h]
\centering 
\includegraphics[width=0.45\textwidth]{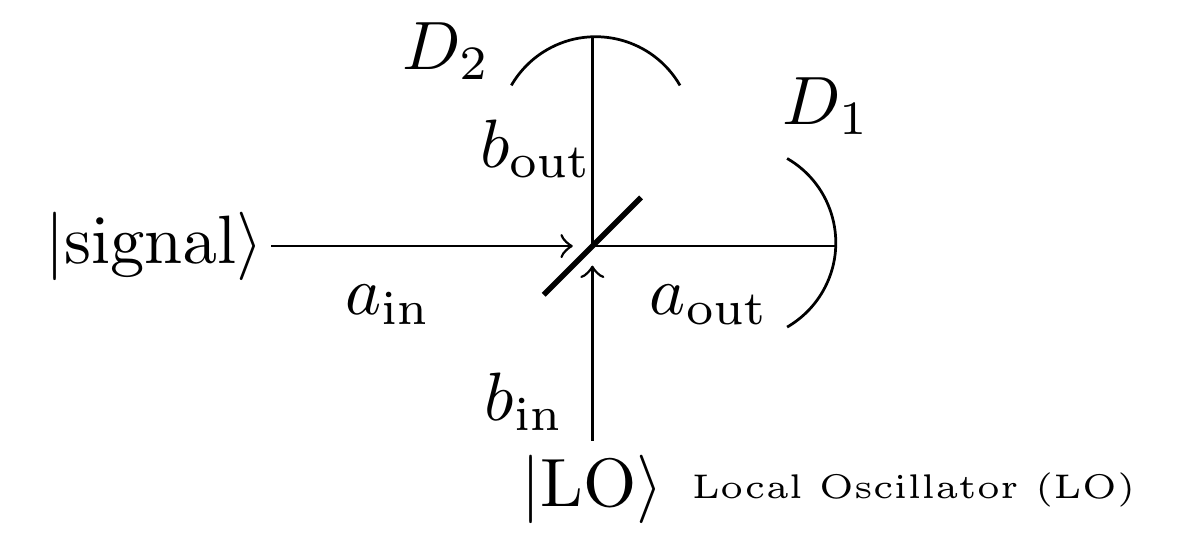}
\caption{Balanced homodyne setup. An arbitrary signal state $\ket{{\rm signal }}$ is mixed on a 50:50 beam-splitter with a local oscillator. The two outputs are measured by detectors ($D_1$, $D_2$) that produce currents $I_i$ proportional to the intensity of the field. The homodyne measurement result is proportional to the difference of these currents i.e. $I_2- I_1$ and the sum current $I_2+I_1$ does not contain useful information. In standard quantum homodyne detection the local oscillator is a large amplitude coherent state i.e.  $\ket{\rm LO} = \ket{\beta} $.  The phase of the coherent state, $\theta$ in $\beta = |\beta| e^{i\theta}$, determines the measured quadrature. In this work we consider local oscillators prepared in superpositions of coherent states i.e. $\ket{\rm LO} \propto \ket{\beta} \pm \ket{-\beta} $. 
}\label{fig:homodynefig}
\end{figure}

Balanced homodyne detection is performed by mixing an arbitrary input signal state $\ket{\Psi}$, with a prepared reference state or {\em local oscillator} (LO) on a 50:50 beam-splitter, see \cref{fig:homodynefig}. The two outputs of this beam-splitter are then measured by detectors that produce currents that are proportional to the intensity of the field. The difference between the two currents is the output signal and effectively measures a quadrature of electromagnetic field~\cite{GrandiParis2017}.  The output is considered to be destructively measured, that is all energy contained within the field is fully absorbed in the act of measurement. 

In the quantum analysis of homodyne measurements all elements of of the scheme (fields, beamsplitters, and detectors) must be treated as quantum objects. The goal of the analysis is to predict the statistics of the measurement. Many quantum treatments of homodyne detection  \cite{Walker87,Collett87,Barchielli_1990,Braunstein90,Grabowski1992,WodkKocha97,homodynedetect, Kiukas08} calculate moments of the detectors (or output signal) and show this limits to moments of a quadrature variable. Another approach, taken by Tyc and Sanders~\cite{Tyc_2004,Tyc_2003}, is to calculate the Kraus operators and positive operator valued measure (POVM) and show that these limit to the POVM for an ideal quadrature measurement.

Little consideration to date has been given to states of the LO that are not effectively classical (or non-Gaussian). However there has been related work that has considered variations of standard homodyne measurement.  In Refs.~\cite{SandLeeKim95,Sanders96} Sanders \emph{et al.} considered homodyne detection using a squeezed LO. Recent work by \citet{Thekkadath2019} shows that using the setup in \cref{fig:homodynefig} one can project onto an even parity states by using a "reversed" quantum interference argument, a control state and postselecting on an equal number of quanta at the output detectors.  The control state determines the even parity state detected and does not necessarily play a role like a local oscillator.  Another recent work by  \citet{Thekkadath2019a} shows a calculation of the properties of homodyne detection with local oscillators that are coherent states with strengths down to zero, i.e. a weak field local oscillator~\cite{Weak_LO_1995}.   This shows that one can smoothly transition from photon counting style detections to field quadrature variables. Related issues have been examined by Olivares et al. in Refs.~\cite{Olivares_2019,Olivares2020}.  However in all these cases, the local oscillator here is always a coherent state which is generally considered to have classical properties.

In this work we consider using superpositions of coherent states as local oscillators and derive the corresponding Kraus operators and POVM elements in the strong LO limit.
In \cref{sec:homodyne} we give an alternative derivation of the results of Tyc and Sanders~\cite{Tyc_2004,Tyc_2003}. We show through application of the algebra of creation and annihilation operators for bosonic fields, that a coherent state LO $\ket{\beta}$, where $\beta = |\beta|e^{i\theta}$, results in the measurement of an arbitrary quadrature projector $|x_\theta\rangle \langle x_\theta|$, in the limit where $|\beta|\rightarrow\infty$.
In \cref{sec:cat} we use this alternative derivation to consider a LO that is proportional to $\ket{\beta}\pm \ket{-\beta}$. For this local oscillator there are two interesting cases: one where we look at both the sum and difference photo-current and one where we only look at the difference photo-current. In the first case the POVM is $\propto \op{x_\theta}{x_\theta} \pm \op{-x_\theta}{x_\theta}\pm  \op{x_\theta}{-x_\theta} +  \op{-x_\theta}{-x_\theta} $. Here the quadrature is set by $\theta$ of the LO and quadrature outcome is readout via the difference current. The parity  $\pm$ is determined by the sum current. In the second case POVM is the reflection symmetric measurement of an arbitrary quadrature, i.e. $\propto |x_\theta\rangle \langle x_\theta| + |-x_\theta\rangle \langle -x_\theta| $. In \cref{sec:examples} we give numerical examples of the statistics of these measurements in the moderate local oscillator limit. Then in \cref{sec:remotecatprep} we show how the non-classical measurements can be used to prepare a non-classical state of a remote system using only a EPR state. Finally we conclude in \cref{sec:conc}.

% ================================================================
\section{Homodyne measurement with coherent state local oscillator}\label{sec:homodyne}
% ================================================================
In this section, we re-derive the Kraus operators and POVM elements for a standard homodyne measurement. Our method is inspired by the work of Tyc and Sanders~\cite{Tyc_2004,Tyc_2003} and later \citet{Puentes2009}.  However, we use different techniques and variables that are better suited for the later consideration of non-classical local oscillator states.

\subsection{Exact Kraus operator}
We will now construct the Kraus operator for \cref{fig:homodynefig} by working backwards from the detectors towards the states.

A departure from the usual treatment of Homodyne measure in the method of Tyc and Sanders is to model the measurement of intensity by ideal photon number resolving detection at detectors 1 and 2.\footnote{This idea has seen further investigation see e.g. Refs.~\cite{Puentes2009,Lipfert2015,Thekkadath2019,Olivares_2019}.} Whereas balanced Homodyne measurement typically involves the light impinging on detectors that respond to intensity. Both treatments consider the photocurrents $I$  produced by the detectors to be proportional to the number operator e.g. $I \propto \expt{a^\dagger a}$ and the measurement result is the difference of the photocurrents i.e. $I_1 - I_2 \propto \expt{a^\dagger a} - \expt{b^\dagger b}$.

Typically the number resolving measurements are modelled by Kraus operators that are projectors onto a Fock state basis, e.g. $\Pi_n = |n \rangle \langle n|$, which might represent a quantum non-demolition detection of photon number. In virtually all cases, optical detectors completely absorb the field and hence the field state after the measurement is mapped to vacuum for any measurement outcomes.  We could introduce operators to denote this case $P_n = |0\rangle\langle n|$. However, we are never going to be interested in this conditional state and hence consider this detection to be a ``partial projection'' $P_n =  \langle n|$, which effectively traces out the post-measurement state.

The object that precedes the detectors is a 50:50 beamsplitter and we denote the unitary representing this object as $U_{\rm BS}$. The Heisenberg evolution of the annihilation operators due to $U_{\rm BS}$ is
\begin{subequations}\label{eq:BS}
	\begin{eqnarray}
	a_{\rm out} :=	U_{\rm BS}^\dagger\, a_{\rm in}\, U_{\rm BS} &=& \frac{1}{\sqrt{2}} (a_{\rm in}+b_{\rm in}) \\
	b_{\rm out}:=	U_{\rm BS}^\dagger\, b_{\rm in}\, U_{\rm BS} &=& \frac{1}{\sqrt{2}} (a_{\rm in}-b_{\rm in}) 
	\end{eqnarray}
\end{subequations}
where $a_{\rm in}$ is an annihilator operator for the signal mode and $b_{\rm in}$ is an annihilation operator for the LO mode.

The next step is to include the input states. By introducing the input states we may define a Kraus operator  $M_{n,m}^{[\beta]}$ that acts on the input Hilbert space of mode $a_{\rm }$ 
\begin{equation}
M_{n,m}^{[\beta]}\ket{{\rm signal}} = (P_n \otimes P_m) U_{\rm BS}(\ket{{\rm signal}} \otimes \ket{ {\rm LO}}),
\end{equation}
the superscript $\beta$ is in anticipation of taking a ``large local oscillator limit'' using a parameter $\beta$. By itself the Kraus operator is
\begin{equation}\label{eq:kraus_simplified}
M_{n,m}^{[\beta]} 
= \bra{n} \bra{m} U_{\rm BS}  \ket{ {\rm LO}}
\equiv P_n \otimes P_m U_{\rm BS}(I \otimes \ket{ {\rm LO}}). 
\end{equation}
This equation represents the Kraus operator for measuring the detection event for $n$ and $m$ photons. The Fock basis dual vectors within the Kraus operator $M_{n,m}^{[\beta]}$ of~\cref{eq:kraus_simplified} can be written in terms of annihilation operators acting on a vacuum tensor product space to give,
\begin{equation}
    M_{n,m}^{[\beta]} = \bra{0}\bra{0} \frac{(a_{\rm out})^n}{\sqrt{n!}}\frac{(b_{\rm out})^m}{\sqrt{m!}} U_{\rm BS} \ket{ {\rm LO}}.
\end{equation}
 It is not yet apparent that this is an operator on the input Hilbert space so we perform further manipulations to elucidate this fact.
 
 We transform $a_{\rm out}$ and $b_{\rm out}$ to linear combination of the input operators by inserting identity $U_{\rm BS}U_{\rm BS}^\dagger$ many times between the powers of the annihilation operators and substituting~\cref{eq:BS}; we also use the fact that $\bra{0} \!\bra{0} U_{BS} = \bra{0} \!\bra{0}$. Doing so gives a Kraus operator 
\begin{align}\label{eq:exact_kraus_arbitraryLO}
	M_{n,m}^{[\beta]} 
		&=\!	\bra{0} \!\bra{0} \! \frac{1}{\sqrt{n!m!}} 
		U_{BS} U_{BS}^\dagger a_{\rm out}^n U_{BS}
			U_{BS}^\dagger b_{\rm out}^m U_{BS}\!  \ket{{\rm LO}}, \nonumber\\	
		&=	\bra{0} \bra{0} \frac{1}{\sqrt{n!m!}} 
			\left(\frac{a_{\rm in}+b_{\rm in}}{\sqrt{2}}\right)^n 
			\left(\frac{a_{\rm in}-b_{\rm in}}{\sqrt{2}}\right)^m
			 \ket{{\rm LO}}. 
	\end{align}	 
At this point the quantum state of the local oscillator $\ket{{\rm LO}}$ is arbitrary and could be replaced with any quantum state.
If we now use the fact that in homodyne measurement the LO is a coherent state, i.e.  $\ket{{\rm LO}}= \ket{\beta}$, on mode $b_{\rm in}$. Then we arrive at the Kraus operator, with no approximations, (See Eq. 2 in Ref.~\cite{Puentes2009})
\begin{align}	
M_{n,m}^{[\beta]} 		& = \bra{0} 
		\frac{(a_{\rm in}+\beta)^n (a_{\rm in}-\beta)^m}
			{2^{(n+m)/2} \sqrt{n!m!}}
			e^{-|\beta|^2/2}. \label{eq:exact_kraus}
\end{align}
This expression is now an operator acting {\em only} on the input signal mode $a_{\rm in}$, and is valid for all values of $\beta$ including small values. From now on we drop the subscript ``in'' on the operator to reduce notational clutter i.e. $a_{\rm in}\mapsto a$.  
 
 The positive operator valued measure (POVM) corresponding to \cref{eq:exact_kraus} is
\begin{equation}
    E_{n,m}^{[\beta]} = \big (M_{n,m}^{[\beta]} \big )\dg M_{n,m}^{[\beta]},
\end{equation}
that this is a valid POVM is evident because $ E_{n,m}^{[\beta]} \ge 0$ and $\sum_{n,m} E_{n,m}^{[\beta]} = \Id$. 

\subsection{Kraus operator approximations}
In the Kraus operators and POVM above have not yet seen the emergence of a quadrature-like measurement result. 
In order to proceed towards our goal of deriving a quadrature measurement from the Kraus operator \cref{eq:exact_kraus} we will need to make some assumptions and approximations. Specifically, we will repeatedly take the large amplitude local oscillator (LO), i.e.  $\betalim $. Naively one would expect that implies the LO intensity is much larger than the signal i.e. $|\beta|^2 \gg \bra{\psi} \hat n \ket{\psi}$, however the approximate condition~\cite[Sec. 4.4]{Tyc_2004} is  $|\beta|^2 \gg \bra{\psi} \hat n^2 \ket{\psi}$.

At this point we will assume $n > m$.  This assumption is not actually a significant restriction due to the symmetry between $n$ and $m$. With this assumption \cref{eq:exact_kraus} can be factored as
\begin{align}
    	M_{n,m}^{[\beta]} =&
	\bra{0} 
		\left(1+\frac{\hat{a}}{\beta}\right)^{n-m} 
		\left(1-\frac{\hat{a}^2}{\beta^2}\right)^m \nonumber\\
	&	\frac{e^{-|\beta|^2/4}}{\sqrt{n!}} 
		\left(\frac{\beta}{\sqrt{2}} \right)^n
		\frac{e^{-|\beta|^2/4}}{\sqrt{m!}} 
		\left(\frac{-\beta}{\sqrt{2}} \right)^m, \label{eq:prog}
\end{align}
as $n-m$ is positive.

Following Tyc and Sanders' logic (see \cref{app:quad}) we change variables representing the measurement output to the difference of the counts   
\begin{align}\label{eq:conv}
x =e^{-i \theta}\Tilde x= \frac{n-m}{\sqrt{2}|\beta|e^{i\theta}}= \frac{n-m}{\sqrt{2}\beta},
\end{align}
which is essentially an estimator for the quadrature component (hence the use of the variable $x$).  Using this new variable we can write \footnote{Regarding the change of variables. Alternatively one can take $\beta^\prime = \sqrt{2} x \beta$, then substitute into the middle equation to get the standard exponential limit.  If $x=0$ it also works.}
\begin{equation}\label{eq:firstterm}
	\left(1+\frac{\hat{a}}{\beta}\right)^{n-m} =
	\left(1+\frac{e^{-i \theta}\hat{a}}{|\beta|}\right)^{\sqrt{2} \tilde x |\beta|}\! 
	\underset{\betalim}{\rightarrow}
	e^{\sqrt{2} e^{-i \theta}\tilde x \hat{a}}.
\end{equation}
In \cref{app:quad} we show the distribution of outcomes $m$ and $n$ will be peaked around $|\beta|^2/2$, due to the Poisson statistics of the LO overwhelming the signal mode. Thus we replace the random variable $m$ with it's mean: $|\beta|^2/2$. Using this approximation on the second bracketed operator term in \cref{eq:prog} we arrive at
\begin{equation}\label{eq:secondterm}
	\left[1-\frac{e^{-2i \theta}\hat{a}^2}{|\beta|^2}\right]^m\! \approx\!
	\left[1-\frac{e^{-2i \theta}\hat{a}^2}{|\beta|^2}\right]^{\frac{|\beta|^2}{2}} \!\!
	\underset{\betalim}{\rightarrow}\!
	e^{-\smallfrac{1}{2} e^{-2i \theta}\hat{a}^2}
\end{equation}
The leading order correction to the above approximation is $O(1/|\beta|^2)$, as shown in \cref{app:error_est}, which tends to zero in the large LO limit.

So far we have assumed that $n>m$. In \cref{app:mgn_cs} we show that when $m > n$ a cancellation of signs occurs in the expressions equivalent to \cref{eq:firstterm} and \cref{eq:secondterm} , which results in the same asymptotic limit. So finally we can approximate the Kraus operator in the strong oscillator limit for all cases as
\begin{align}
	M_{n,m}^{[\beta]} 
	\underset{\betalim}{\approx}&
\bra{0} 
	e^{\sqrt{2}e^{-i\theta} \tilde{x} \hat{a}}
e^{-e^{-2i \theta}\hat{a}^2/2}\nonumber\\
	&	\frac{e^{-|\beta|^2/4}}{\sqrt{n!}} 
		\left(\frac{\beta}{\sqrt{2}} \right)^n
		\frac{e^{-|\beta|^2/4}}{\sqrt{m!}} 
		\left(\frac{-\beta}{\sqrt{2}} \right)^m.
\end{align}
This Kraus operator is the basic result we use for our remaining analysis.

The exponentials of $\hat a$ acting on vacuum in the Kraus operator $M_{n,m}^{[\beta]}$ in the limit of large $|\beta|$ are a good approximation of a quadrature eigenstate~\cite{SotoMoya2013}.  To see this, we can write, with the choice of units used here, eigenstates of an arbitrary quadrature $Q(\varphi)=(e^{-i\varphi}a+ e^{i\varphi}a^\dagger)/\sqrt{2}$ as
\begin{align}\label{eq:quadeig}
\ket{x_\varphi} = 
\frac{e^{-x^2/2}}{\pi^{1/4}} 
e^{\sqrt{2}x \chi a^\dagger}
e^{-\chi^2{a^\dagger}^2/2} 
\ket{0},
\end{align}
with  $\chi = e^{i\varphi}$ and $\ket{0}$ is the vacuum state. (In \cref{app:quadeig} we show that $\ket{x_\varphi}$ is an eigenstate of  $Q(\varphi)$ using techniques adapted from Ref.~\cite{SotoMoya2013}.) This means the approximation of the Kraus operator can be written as 
\begin{align}\label{eq:goodness}
	M_{n,m}^{[\beta]} 
	\underset{\betalim}{\approx}&
	\bra{x_\theta} 
	\pi^{1/4} e^{\tilde{x}^2/2} (-1)^m e^{i \theta(n+m)} \nonumber\\
	&	\frac{e^{-|\beta|^2/4}}{\sqrt{n!}} 
		\left(\frac{|\beta|}{\sqrt{2}} \right)^n
		\frac{e^{-|\beta|^2/4}}{\sqrt{m!}} 
		\left(\frac{|\beta|}{\sqrt{2}} \right)^m.
\end{align}
with some slight mixing of notation as $x$ is used to represent a quantity proportional to the difference of $n$ and $m$ as defined above.  The final two terms are the square root of a Poisson distribution. They can be approximated in the strong local oscillator limit, $|\beta| \rightarrow \infty$ using the continuous approximation for the Poisson distribution, that is
\begin{equation}
	\left [ \frac{e^{-|\beta|^2/2}}{n!}\right ]^{\half} 
	\left[\frac{|\beta|^2}{2} \right]^{\tfrac{n}{2}}
	\approx
	\frac{e^{-(n-|\beta|^2/2)^2/(2 |\beta|^2)}}{(\pi |\beta|^2)^{1/4}} \sqrt{dn}.
\end{equation}
Notice we have introduced the square root of the probability measure of $n$.  This is because the continuous approximation smooths the differences between the values of $n$ which were implicitly $1$ in the discrete distribution.  This argument on the differential $dn$ applies to the probability, but this expression involves the probability \emph{amplitude} which is why the square root of this measure is required~\cite{sqrt_prob_measure}.  Using this approximation on the final two terms in \cref{eq:goodness} gives two independent normal distributions for $n$ and $m$ with mean and variance $|\beta|^2/2$.  However, we wish to resolve the results into scaled sum and differences of $n$ and $m$.  Specifically the exponent of the combined distribution for $n$ and $m$ is 
\begin{multline}
 -\frac{1}{2 |\beta|^2}\left[ \left (n-\frac{|\beta|^2}{2}\right)^2 + \left (m-\frac{|\beta|^2}{2}\right)^2 \right ] \\ =
 -\frac{1}{4 |\beta|^2}\left[ \left(n + m-|\beta|^2\right)^2 + (n-m)^2 \right].
\end{multline}
This equation shows that the sum of $n$ and $m$ is normal distribution with mean $|\beta|^2$ and variance $|\beta|^2$ and the difference of $n$ and $m$ is a normal distribution with mean zero and variance $|\beta|^2$.  The difference of $n$ and $m$ can be scaled to be written in terms of $x$ and the distribution in $x$ will be normal with mean zero and variance $1/2$.  The distribution in the scaled $x$ cancels the exponential factor of $e^{\tilde{x}^2/2}$ in Eq.~\ref{eq:goodness} leaving
\begin{align}\label{eq:partialprog}
	M_{n,m}^{[\beta]} \sqrt{dn dm} 
&	\underset{\betalim}{\approx}
     e^{i \theta(n+m)} \bra{x_\theta}   (-1)^m  \nonumber\\
  &  \frac{2^{1/4}}{|\beta|^{1/2}}
	\frac{e^{-(n+m-|\beta|^2)^2/(4|\beta|^2)}}{(2\pi|\beta|^2)^{1/4}}
	\sqrt{dn dm},
\end{align}
where this equation has included the probability measure due to the continuum limit as explained above. Moreover we have pulled out one of the global phase factors $e^{i \theta(n+m)}$ in anticipation of similar factors in the cat state local oscillator examined in \cref{sec:cat}.

The change of variables can be simplified by introducing $w$ defined as
\begin{equation}\label{eq:w_sum_var}
w = \frac{n+m}{\sqrt{2}|\beta|}   
\end{equation}
which will be distributed normally with mean $|\beta|/\sqrt{2}$ and variance $1/2$. To complete the change of variables, the probability measure must be changed, so we need the determinant of the Jacobian
\begin{equation}
    dx dw = \left|\frac{\partial(x,w)}{\partial(n,m)}\right| dn dm = \frac{1}{|\beta|^2} dn dm.
\end{equation}
With these changes of variables the final approximation for the Kraus operator which can be written as (including changing the $m$ and $n$ subscripts to $x$ and $w$ as there are no more uses of $m$ and $n$)
\begin{align}
	M_{x,w}^{[\beta]} \sqrt{dw dx}
	\underset{\betalim}{\approx}& \, e^{i\sqrt{2}|\beta| w \theta}
	\bra{x_\theta}  e^{i \pi \sqrt{2}|\beta|(w-x)} \nonumber\\
&	\frac{1}{ |\beta|}
	\frac{e^{-(w-|\beta|/\sqrt{2})^2/2}}{\pi^{1/4}} |\beta| \sqrt{dw dx}.
\end{align}
From the original definitions of $\sqrt{2}|\beta|x$ and $\sqrt{2}|\beta|w$ in terms of the discrete variables, these values are actually integers.  This means that the first phase factor is either $1$ or $-1$ (though these phase factors disappear in the next step of our derivation). In \cref{sec:cat} this phase has an important role to play.

The Kraus operator is important for determining the post measurement state, however the measurement statistics are entirely determined by the POVM.
The POVM for a scaled difference measurement of $x$ will then be
\begin{subequations}
\begin{align}
  dx\, dw\, {E}_{x,w}^{[\beta]} &= 
	dx\, dw\, \big (M^{[\beta]}_{x,w})^\dagger M_{x,w}^{[\beta]} \\
	& = dx\, dw\,
	\frac{e^{-(w-|\beta|/\sqrt{2})^2}}{\sqrt{\pi}} \ket{x_\theta}\bra{x_\theta}
\end{align}
\end{subequations}
 which agrees with equation (19) in Ref.~\cite{Tyc_2004} when differing notation is taken into account.  

In the standard approach to homodyne detection, the $w$ variable, the exact sum of the photons counted, is unobserved.  We will integrate over it and take the final large oscillator limit i.e. 
\begin{align}
	dx\, E_{x} &	\underset{\betalim}{\approx} dx \int_0^\infty dw \, E_{x,w}^{[\beta]}\,,
\end{align}
where $x$ is the measurement outcome. 
Thus POVM with outcome $x$ in homodyne detection of an arbitrary quadrature $\theta$ is
\begin{subequations}
\begin{align}
	dx\, E_{x} &	\underset{\betalim}{\approx}
	dx \int_0^\infty dw \frac{e^{-(w-|\beta|)^2}}{\sqrt{\pi}} \op{x_\theta}{x_\theta} \nonumber\\
&	\underset{\betalim}{\approx}\frac{1}{2} \left[\text{erf}\left(\frac{|\beta|}{\sqrt{2}}\right)+1\right] dx \op{x_\theta}{x_\theta}\\
&=dx \op{x_\theta}{x_\theta},
\end{align}
\end{subequations}
which is a well-defined POVM as all elements are positive (projectors) and 
\begin{equation}
	\int dx E_{x}  = \int dx \op{x_\theta}{x_\theta} = \Id.
\end{equation}

% ================================================================
\section{Homodyne measurement with a cat state local oscillator}\label{sec:cat}
% ================================================================
In this section we derive the Kraus operators and POVM for the case of a local oscillator state that is a superposition of two coherent states, i.e. a cat state. Broadly the derivation here follows the procedures developed \cref{sec:homodyne}, i.e. switching to sum and difference variables and taking a large local oscillator limit.

The measurement we describe here is inspired by the homodyne measurement paradigm but deviates from it in a number of ways that we now detail so the reader can follow along with this knowledge.
First, a large LO limit might actually be undesirable for a number of reasons. It turns out that in the large LO limit quantum coherence is removed from the measurement outcomes and it is practically hard to make a large cat state and do number resolved detection for many photons. Second, integrating over the sum variable also washes out quantum coherence in the measurement described below.
Third,  number resolved detection seems to play a vital role in the measurement below and in particular the parity of the sum variable $w$. In spite of these issues, we proceed so that we may arrive at an analytical expression for the measurement operators.

We consider and LO that is a superposition of coherent states of the form 
\begin{equation}\label{eq:catstate}
	\N_\pm(\beta)^{-1} \left( \ket{\beta} \pm \ket{-\beta} \right),
\end{equation}
where $\N_\pm(\beta) = \sqrt{2(1\pm e^{-2|\beta|^2})}$. Note that as $|\beta|\rightarrow \infty$ the normalization limits to  $\N_\pm(\beta) \rightarrow \sqrt{2}$.  The plus superposition consists of only even Fock basis terms as the odd amplitudes follow the sign of the $\pm \beta$ amplitude and cancel to zero.  A similar argument follow for the minus superposition but only odd terms survive.  Therefore we say the plus superposition is an even parity state the minus superposition an odd parity state.

The {\em exact} Kraus operator, substituting \cref{eq:catstate} into \cref{eq:exact_kraus_arbitraryLO} instead of $\ket{\beta}$, for the oscillator in a superposition of coherent states is
\begin{align}\label{eq:exactcat_Kraus}
M_{n,m}^{[\beta]_\pm} =&
\bra{0} 
\frac{ e^{-|\beta|^2/2} } {2^{(n+m)/2} \sqrt{n!m!} \N_\pm(\beta)}\nonumber \\
&\left( (\hat{a}+\beta)^n (\hat{a}-\beta)^m \pm (\hat{a}-\beta)^n (\hat{a}+\beta)^m\right),
\end{align}
where the subscript of $\pm$ on $[\beta]_\pm$ denotes the plus or minus superposition. We assume $n>m$, as we did in deriving \cref{eq:prog}, and get 
\begin{align}\label{eq:progcat}
&M_{n,m}^{[\beta]_\pm} =
\bra{0} 
\frac{ e^{-|\beta|^2/2} } {2^{(n+m)/2} \sqrt{n!m!} \N_\pm(\beta)}
\left(1-\frac{\hat{a}^2}{\beta^2}\right)^m\nonumber \\
&\left[ \beta^n(-\beta)^m 	\left(1+\frac{\hat{a}}{\beta}\right)^{n-m} \pm 
(-\beta)^n \beta^m 	\left(1-\frac{\hat{a}}{\beta}\right)^{n-m}\right].
\end{align}
As shown in appendix~\ref{app:mgn_sup}, the expression for $m>n$ is the same as this expression but with the sign of the superposition possibly changed depending on the parity of $m-n$. The up shot is: \cref{eq:progcat} covers the case of $m > n$ if the information as to which superposition phase applies is incorporated.

It turns out in the large LO limit $m$ is distributed as a Poisson distribution with parameter  $\lambda =|\beta|^2/2$ as before, see \cref{app:catclick} for the details. Thus the reasoning around \cref{eq:secondterm} also applies to \cref{eq:progcat}. We also perform the change of variables given in \cref{{eq:conv}} and take the limits given in \cref{eq:firstterm,eq:secondterm} and make the replacement from the large $|\beta|$ limit i.e.  $  \N_\pm(\beta) \rightarrow \sqrt{2}$. 

Using these approximations we have
\begin{align}
M_{n,m}^{[\beta]_\pm} 	\underset{\betalim}{\approx}
\bra{0}\! &
\left[(-1)^m e^{\sqrt{2}e^{-i\theta} \tilde{x} \hat{a}}
e^{-e^{-2i \theta}\hat{a}^2/2} \right . \nonumber\\
&\left . \! \pm 
(-1)^n e^{-\sqrt{2}e^{-i\theta} \tilde{x} \hat{a}}
e^{-e^{-2i \theta}\hat{a}^2/2} \right]\frac{1 } { \sqrt{2}}\nonumber \\
&\frac{ e^{-|\beta|^2/4} } {\sqrt{n!}}\left (\frac{\beta}{\sqrt{2}}\right)^n
 \frac{ e^{-|\beta|^2/4} } {\sqrt{m!}}\left (\frac{\beta}{\sqrt{2}}\right)^m .
\end{align}
At this point we recognise a quadrature eigenstate and a ``$\pi$'' rotated quadrature eigenstate,  as per \cref{eq:quadeig}, and use that to further simplify the operator to
\begin{align}
M_{n,m}^{[\beta]_\pm} 	\underset{\betalim}{\approx} 
& \pi^{1/4} e^{\tilde{x}^2/2} \frac{ e^{-|\beta|^2/4} } {\sqrt{n!}}\left (\frac{|\beta|}{\sqrt{2}}\right)^n \frac{ e^{-|\beta|^2/4} } {\sqrt{m!} }\left (\frac{|\beta|}{\sqrt{2}}\right)^m \nonumber \\
 &\frac{e^{i \theta(n+m)} } { \sqrt{2}}  \left[
\bra{x_\theta}  (-1)^m
\pm \bra{-x_\theta}  (-1)^n \right].
\end{align}
Next we approximate the square root of Poisson distributions by normal distributions. The mean and variance of the sum and differences scale in the same way as the previous section giving
\begin{multline}\label{eq:cat_kraus}
M_{n,m}^{[\beta]_\pm} \sqrt{dn dm} 
	\underset{\betalim}{\approx} \\
 \frac{2^{1/4}}{|\beta|^{1/2}}
	\frac{e^{-(n+m-|\beta|^2)^2/(4|\beta|^2)}}{(2\pi|\beta|^2)^{1/4}}
	\sqrt{dn dm}	\\
\frac{  e^{i \theta(n+m)} } { \sqrt{2}}
\left[
\bra{x_\theta}  (-1)^m 
\pm 
\bra{-x_\theta}(-1)^n \right]\, ,
\end{multline}
which should be compared to \cref{eq:partialprog}. Notice there is an overall phase of $e^{i \theta(n+m)}$ as there was in \cref{eq:partialprog}. However now we have a relative phase between quadrature eigenstates which is evident in the terms $ (-1)^m$ and $(-1)^n$. 

Let's pause to consider the implications of these phase factors. For the $+$ cat LO, the Kraus operator is $M_{n,m}^{[\beta]_+}\propto  \bra{x_\theta}  (-1)^m \pm \bra{-x_\theta}(-1)^n$ with $m$ and $n$ taking integer values and the phase factors reflect whether $m$ and $n$ are odd or even. Thus the four possible combinations of the $m$ \& $n$ and dependant phases result in two distinct Kraus operators (upto a global phase), namely
\begin{equation}
\label{eq:measop_nm_position}
M_{n,m}^{[\beta]_+}\propto 
\begin{cases}
\bra{x_\theta}  + \bra{-x_\theta} & n+m \textrm{ even}  \\
\bra{x_\theta}  - \bra{-x_\theta} & n+m \textrm{ odd}
\end{cases}.
\end{equation}
These measurement operators will project onto states of definite parity. To see this fact we specialize to the position quadrature ($\theta=0$) and recall the parity operator can be represented as
\begin{equation}
    P = (-1)^{a\dg a} = \int dx' \, \op{-x'}{x'}\, .
\end{equation} 
One can show that 
\begin{align}
P \ket{x_\pm} &= (\pm 1)\ket{x_\pm}  
\end{align}
where these eigenstates of the parity operator are
\begin{align}
\ket{x_\pm} &\propto  \ket{x }  \pm \ket{-x}  \,,  
\end{align}
 and the proportionality is due to the fact that $\ket{x}$ is non-normalizable. A similar argument can be made for a LO using the $-$ cat state but with the signs of the plus and minus on the right-hand side of~\cref{eq:measop_nm_position} exchanged. That is to say, in this situation, the measurement is a parity measurement.

Returning to the derivation, we now complete the change of variables using the previously defined variable $w$, see \cref{eq:w_sum_var}. From \cref{eq:cat_kraus} we can see that $w$ is still distributed normally with mean $|\beta|/\sqrt{2}$ and variance $1/2$. This is because both $n$ and $m$ are approximately Poisson distributed with parameter  $\lambda =|\beta|^2/2$ as detailed in  \cref{app:catclick}.
This gives the Kraus operator
\begin{align}\label{eq:cat_kraus_almostdone}
M_{x,w}^{[\beta]_\pm} \sqrt{dw dx}	\underset{\betalim}{\approx}
\frac{e^{i\sqrt{2}|\beta| w \theta} } { \sqrt{2}}& \left[
\bra{x_\theta} e^{i \pi \sqrt{2}|\beta|(x-w)/2} 
\pm \right. \nonumber\\
&\left.\bra{-x_\theta}e^{i \pi \sqrt{2}|\beta|(x+w)/2} \right]\nonumber\\	&\frac{e^{-(w-|\beta|/\sqrt{2})^2/2}}{\pi^{1/4}} \sqrt{dw dx}\, .
\end{align}
For the detection process, we care primarily about the POVM which is 
\begin{align}
dw\, dx\, E_{x,w}^{[\beta]_\pm }\underset{\betalim}{\approx}&
dw\, dx\, \big ( M_{x,w}^{[\beta]_\pm }\big ) ^\dagger     M_{x,w}^{[\beta]_\pm } .
\end{align}
Substituting our expressions in we find
\begin{align}
dw\, dx\, E_{x,w}^{[\beta]_\pm }\underset{\betalim}{\approx}&
dw\, dx\, \frac{e^{-(w-|\beta|/\sqrt{2})^2}}{\sqrt{\pi}}
\frac{1 } {2}\nonumber\\
&\Big[\,\op{x_\theta}{x_\theta}  \nonumber\\
&\pm \op{x_\theta}{-x_\theta} e^{\sqrt{2}i \pi |\beta| w } \nonumber\\
& \pm \op{-x_\theta}{x_\theta} e^{-\sqrt{2}i \pi |\beta| w} \nonumber\\
&+ \op{-x_\theta}{-x_\theta} \Big] \,.
\label{eq:catKraus_nointegral}
\end{align}
At this point we can still observe a coherence between the $+$ and $-$ outcomes of the measurement, see the terms with the $\pm$ coefficients above. So if knowledge of both the sum and difference variables is retained \cref{eq:catKraus_nointegral} is the final result.

However, if we integrate \cref{eq:catKraus_nointegral} over $w$, presuming that it is unobserved like in homodyne detection, gives an expression of the form
\begin{multline}
dx \int dw E_{x,w}^{[\beta]_\pm  }  = \\
dx \Big[
G(\beta)(\op{x_\theta}{x_\theta} + \op{-x_\theta}{-x_\theta})  \\
\pm I(\beta) (\op{x_\theta}{-x_\theta} + \op{-x_\theta}{x_\theta}) \big]\, ,
\end{multline}
where
\begin{align}
I(\beta)&=\int_0^\infty dw\, e^{\pm 2\sqrt{2}i \pi |\beta| w }  \frac{e^{-(w-|\beta|/\sqrt{2})^2}}{\sqrt{\pi}} \nonumber\\
&= \frac{1}{2} e^{-2 \pi^2 |\beta| ^2}e^{\pm 2 \pi   i|\beta| ^2} \left[1+\text{erf}\left(\frac{(1\pm 2 i \pi ) |\beta| }{\sqrt{2}}\right)\right],   \nonumber\\
G(\beta)&=\int_0^\infty\! dw\, \frac{e^{-(w-|\beta|/\sqrt{2})^2}}{\sqrt{\pi}}=\frac{1}{2} \left[1+\text{erf}\left(\frac{|\beta |}{\sqrt{2}}\right)\right].
\end{align}
Notice that $I(\beta)$ has an overall envelope of $\exp[-2\pi^2 |\beta|^2]$ which very clearly limits to zero as $\betalim$. Thus  in the large LO limit these expressions limit to
\begin{align}
    \lim_{\betalim}I(\beta) =0 \quad {\rm and}\quad
    \lim_{\betalim}G(\beta) =1.
\end{align}
With this the limiting case in the probability is
\begin{align}
dx\, E_x  &\underset{\betalim}{=} dx \int d w\,E_{x,w}^{[\beta]_\pm } \nonumber\\
&=\frac{dx} {2}\Big( \op{x_\theta}{x_\theta}  +  \op{-x_\theta}{-x_\theta} \Big)\label{eq:ans_part1}
\end{align}
which satisfies the normalisation properties of a standard probability density,
\begin{align}
	\int dx \, E_x  = \Id.
\end{align}
This also implies that \cref{eq:catKraus_nointegral} resolved the identity as $dw\, dx\, E_{x,w} = \Id$.

\newcommand{\vacfiguresize}{0.75\columnwidth}
\begin{figure*}[htbp!]
    \centering
    \includegraphics[width=\vacfiguresize]{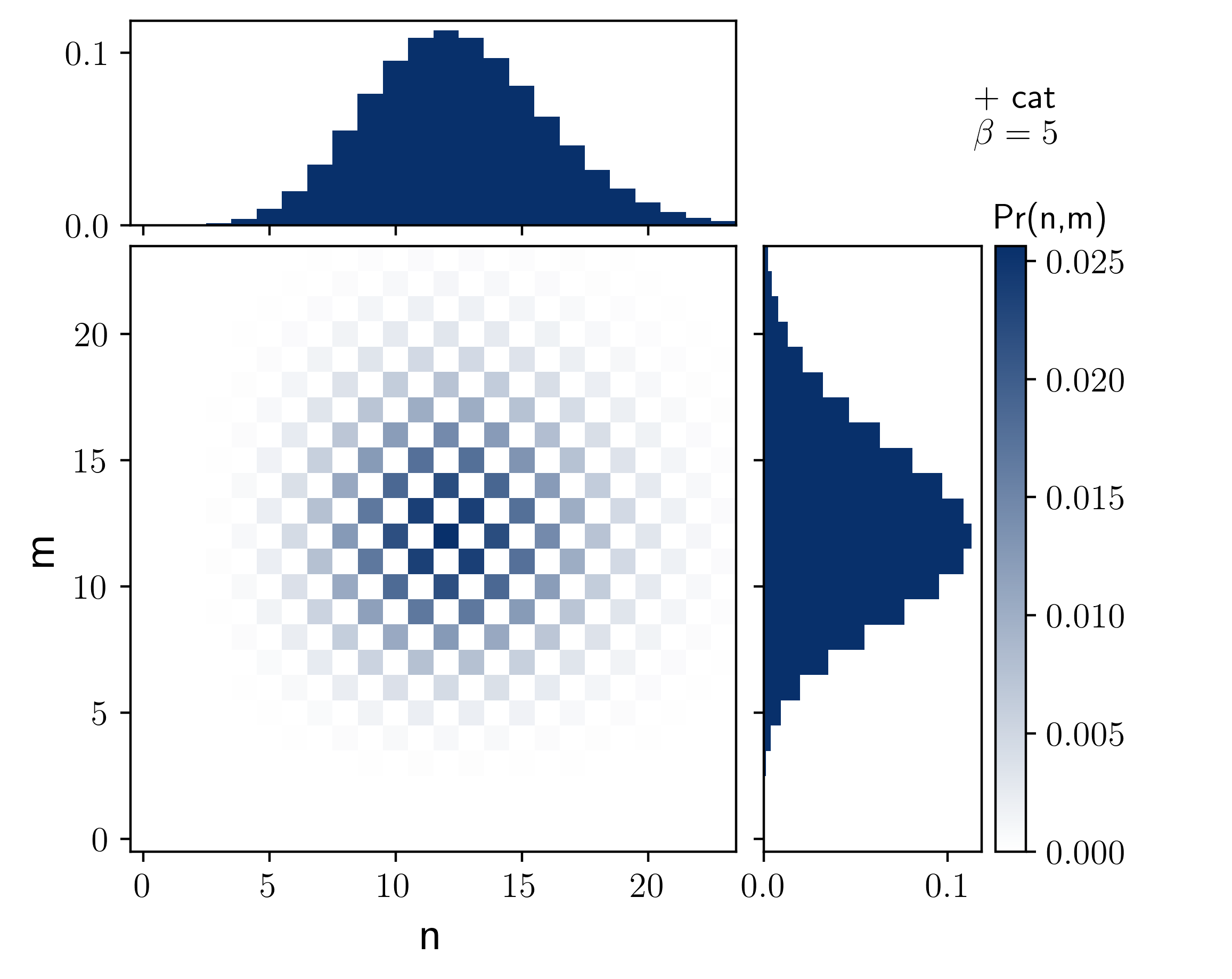}
    \includegraphics[width=\vacfiguresize]{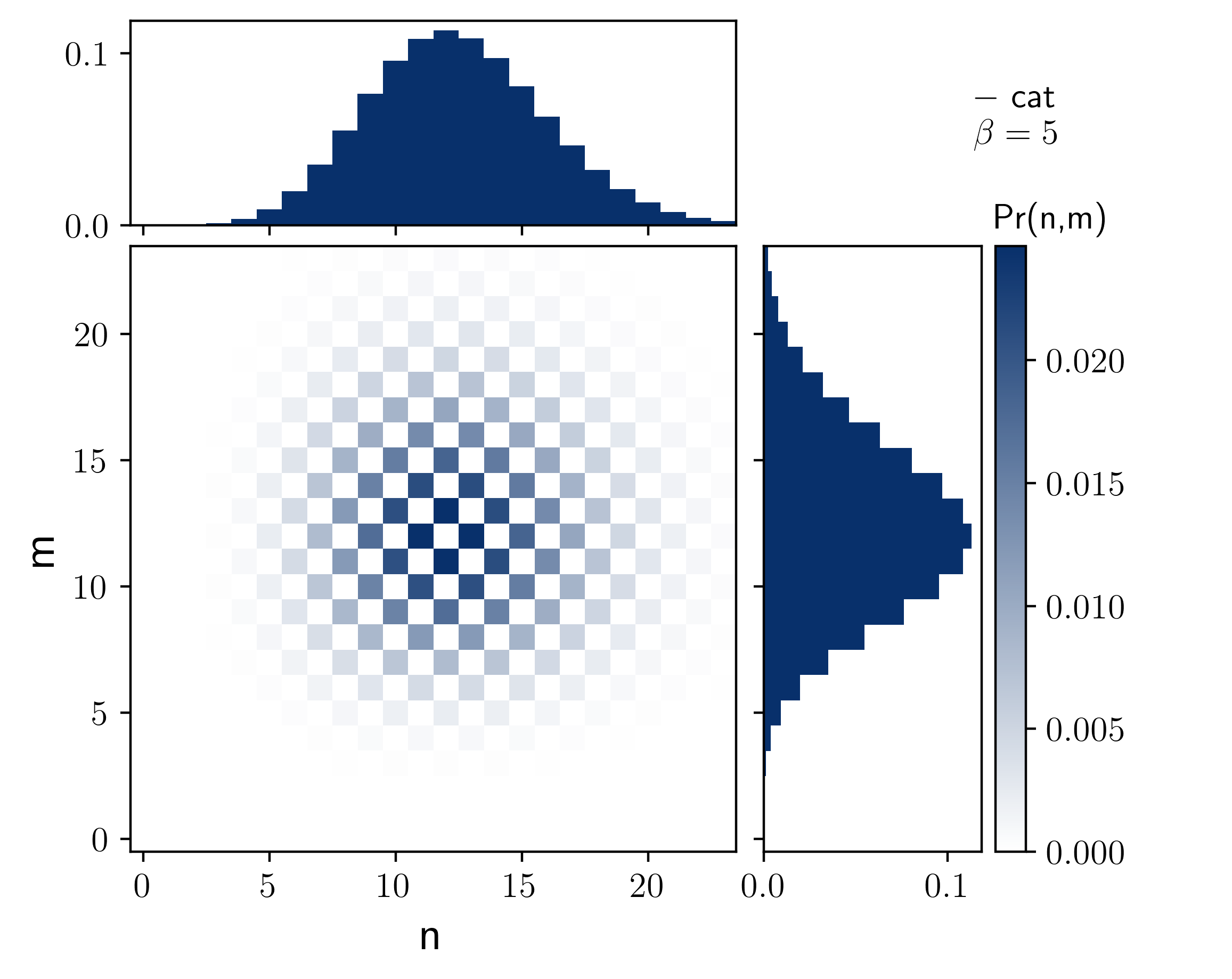}\\
    \includegraphics[width=\vacfiguresize]{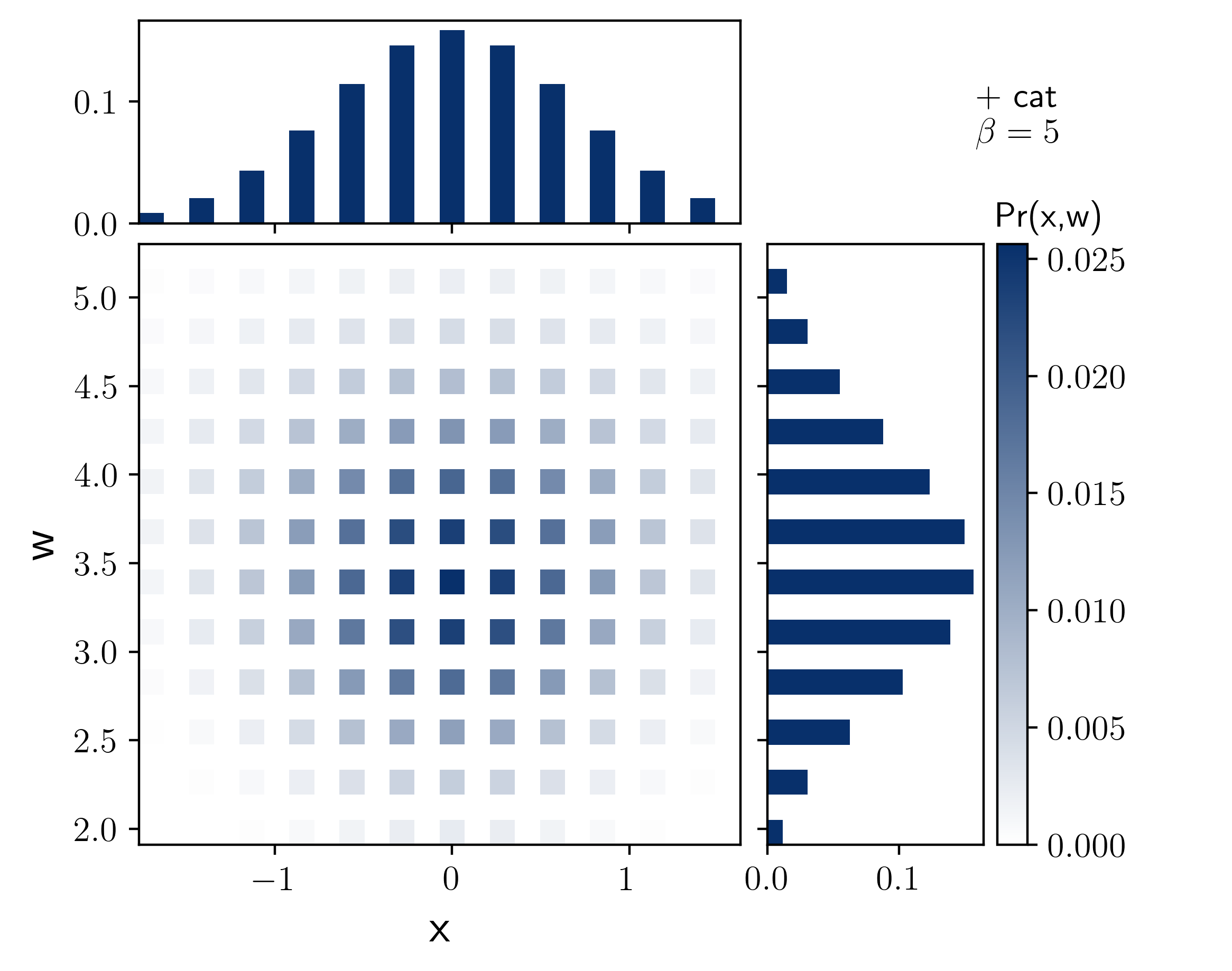} 
    \includegraphics[width=\vacfiguresize]{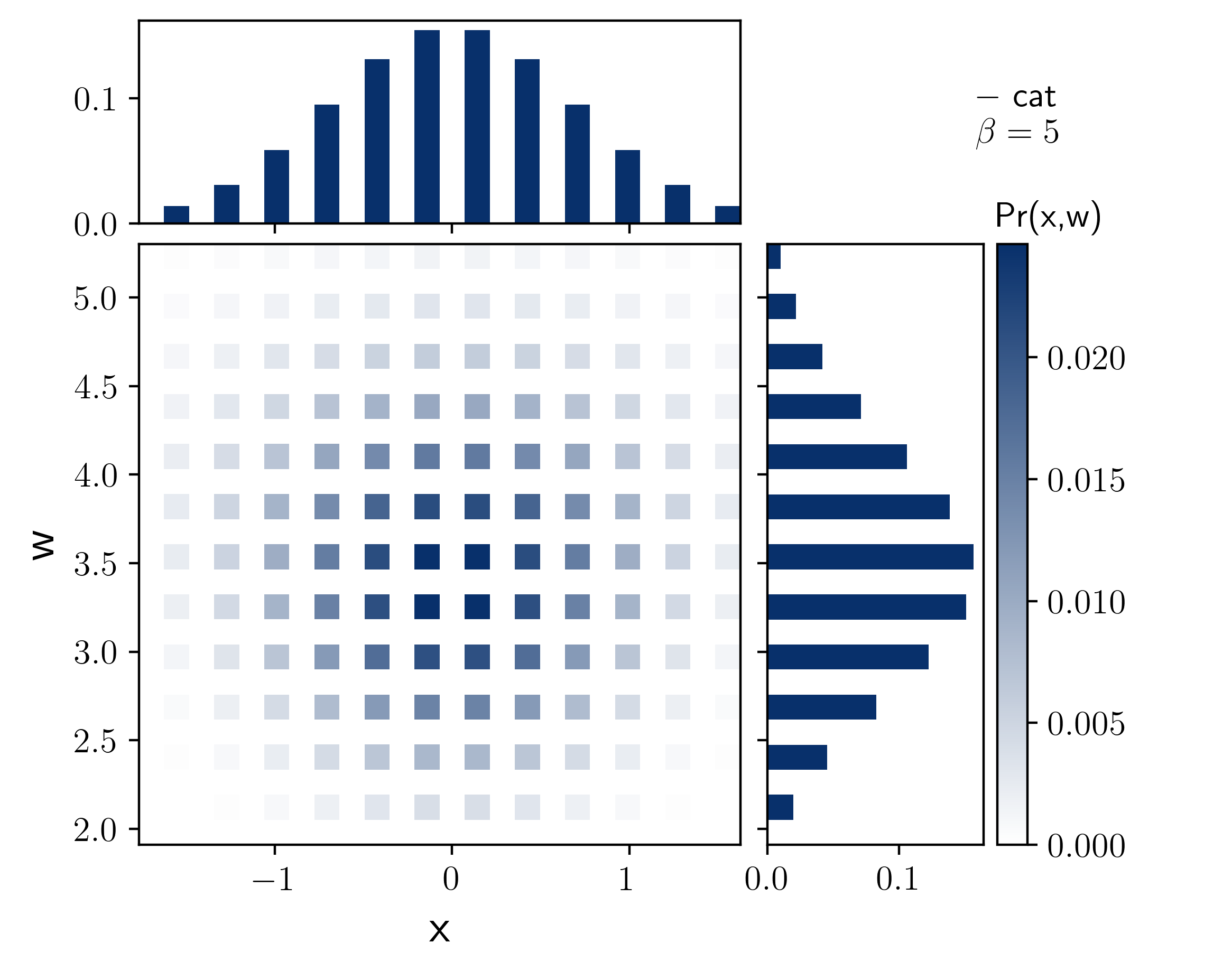} \\
\caption{Click distributions for detecting vacuum (the signal) with a Cat local oscillators where $\beta = 5$ and the expected number of photons in the $b$ mode is $\expt{b\dg b} = |\beta|^2= 25$. (Row 1) Original click distributions $n$ and $m$ {(dimensionless)}, marginal distributions are depicted above and right for both the $+$ and $-$ cat LO. The distribution and marginals are centred around $|\beta|^2/2 = 12.5$. (Row 2) Sum $w=(n+m)/(\sqrt{2}\beta)$ and difference $x=(n-m)/(\sqrt{2}\beta)$ variables {(dimensionless)} and the corresponding marginal distributions for the $+$ and $-$ cat LO. As vacuum is even parity state the parity of the LO is evident in the marginal of the  difference variable. For example the $+$ cat has support on $x=0$ (see the marginal $x$ distribution) while the $-$ cat does not. } \label{fig:vacuum_plots}
\end{figure*}

To summarise so far, using an odd or even cat state as a local oscillator we derived the Kraus operators and POVM for a homodyne like measurement. In the limit of large amplitudes in the cat states, the POVM for each measurement outcome is a sum of a quadrature eigenstate  e.g. $\ket{x_\theta}$ and it's negative $\ket{-x_\theta}$.  This operator projects onto this two-dimensional subspace and hence cannot distinguish between states which equally project onto that subspace.  For example, the states $\ket{x_\theta}$, $\ket{-x_\theta}$ and $\ket{c_\theta} = \frac{1}{\sqrt{2}}(\ket{x_\theta} + \ket{-x_\theta})$ will all give the same probability density. In other words, this measurement is symmetric about reflections through the quadrature origin. For this reason we call it a reflection symmetric measurement.

Note that after integrating out over the sum variable $w$ and taking the large oscillator limit means that the coherence of the cat state in the local oscillator is irrelevant.  So one can see such a measurement as a homodyne measurement with randomly chosen classical phase of either $0$ or $\pi$, which corresponds to a mixed state LO $\propto \op{\beta}{\beta} + \op{-\beta}{ -\beta}.$ 

If we instead consider the POVM before the integration over $w$ was performed, see \cref{eq:catKraus_nointegral}, the coherence terms contain phases of the form $\sqrt{2}i\pi|\beta|w$.  But from the definition of the variable $w$, $\sqrt{2}|\beta|w$ is an integer.  Therefore this phase factor can be treated as $\pm 1$ but only with the knowledge of $w$ whereas without this knowledge and the continuum approximation leads to this phase factor tending quickly to zero.  The ability to use (or ignore) the information contained in the sum variable seems to be a new feature with the cat state local oscillator and provides a means to engineer a measurement.

There is hence a large number of interconnected concerns when taking these approximations together that may be fragile. In the next section we give numerical computations which try to address these concerns with computations involving finite sized local oscillator states.

%==================================================
\section{Example input states}\label{sec:examples}
%==================================================
The above analysis exposes some general properties of homodyne detection with these types of local oscillators, while the case studies below give rise to some more specific information about the details of the measurement without reliance on numerous approximations.  To do this we return to \cref{eq:cat_kraus} to calculate the Kraus operators, but we are guided broadly by the properties uncovered in \cref{sec:cat}.

 \newcommand{\cohfiguresize}{0.67\columnwidth}
\begin{figure*}[htbp]
    \centering
    \includegraphics[width=\cohfiguresize]{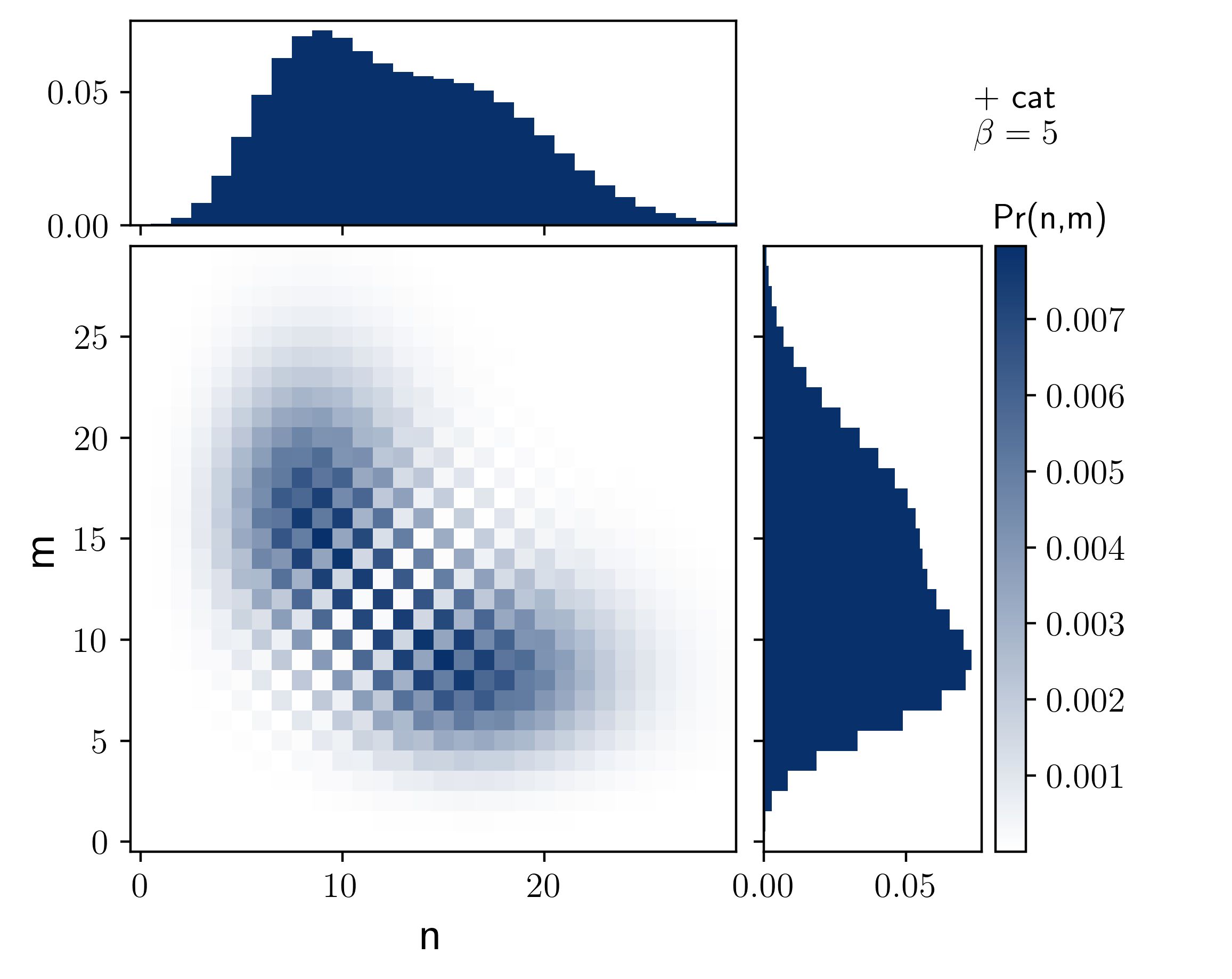}
    \includegraphics[width=\cohfiguresize]{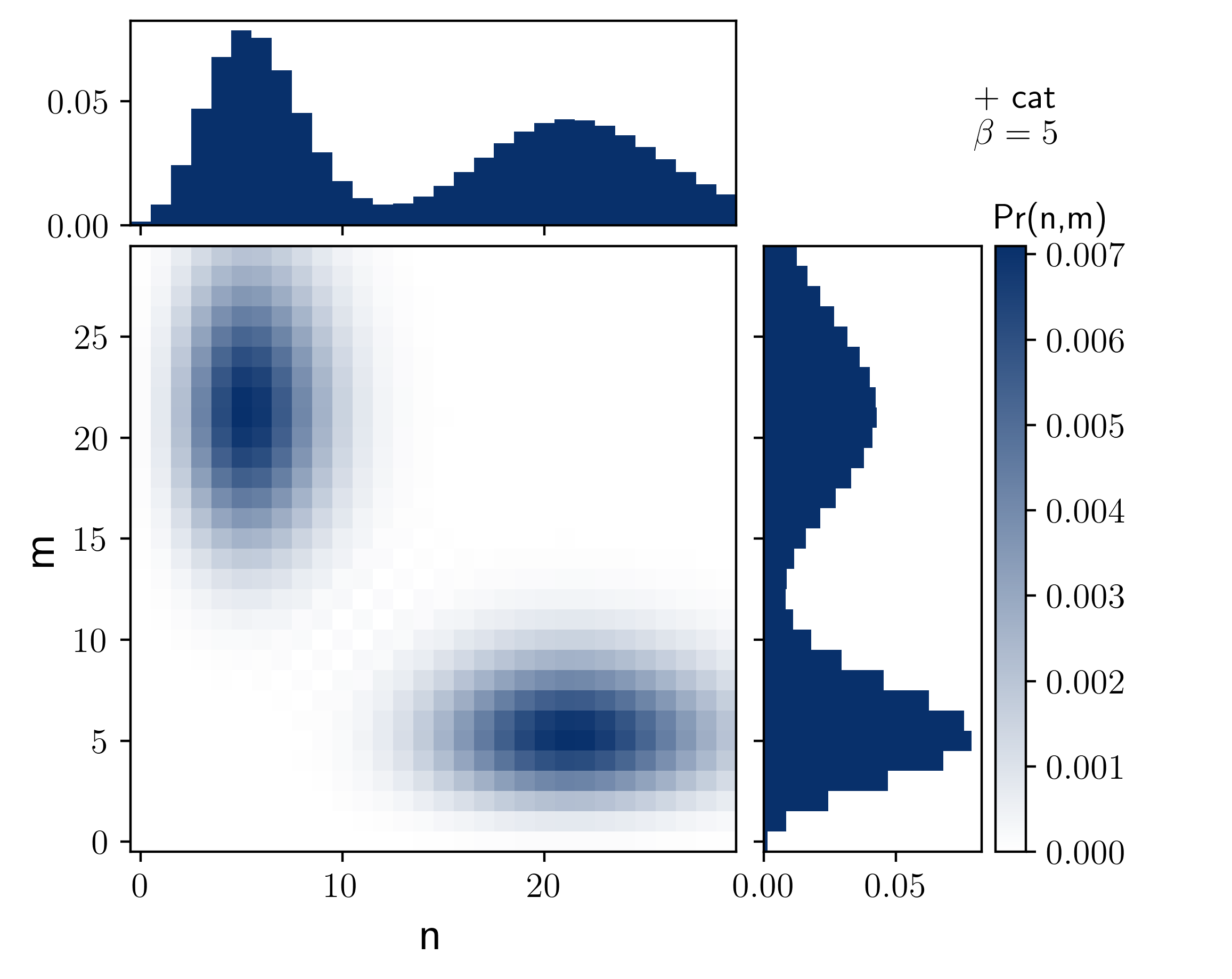}
    \includegraphics[width=\cohfiguresize]{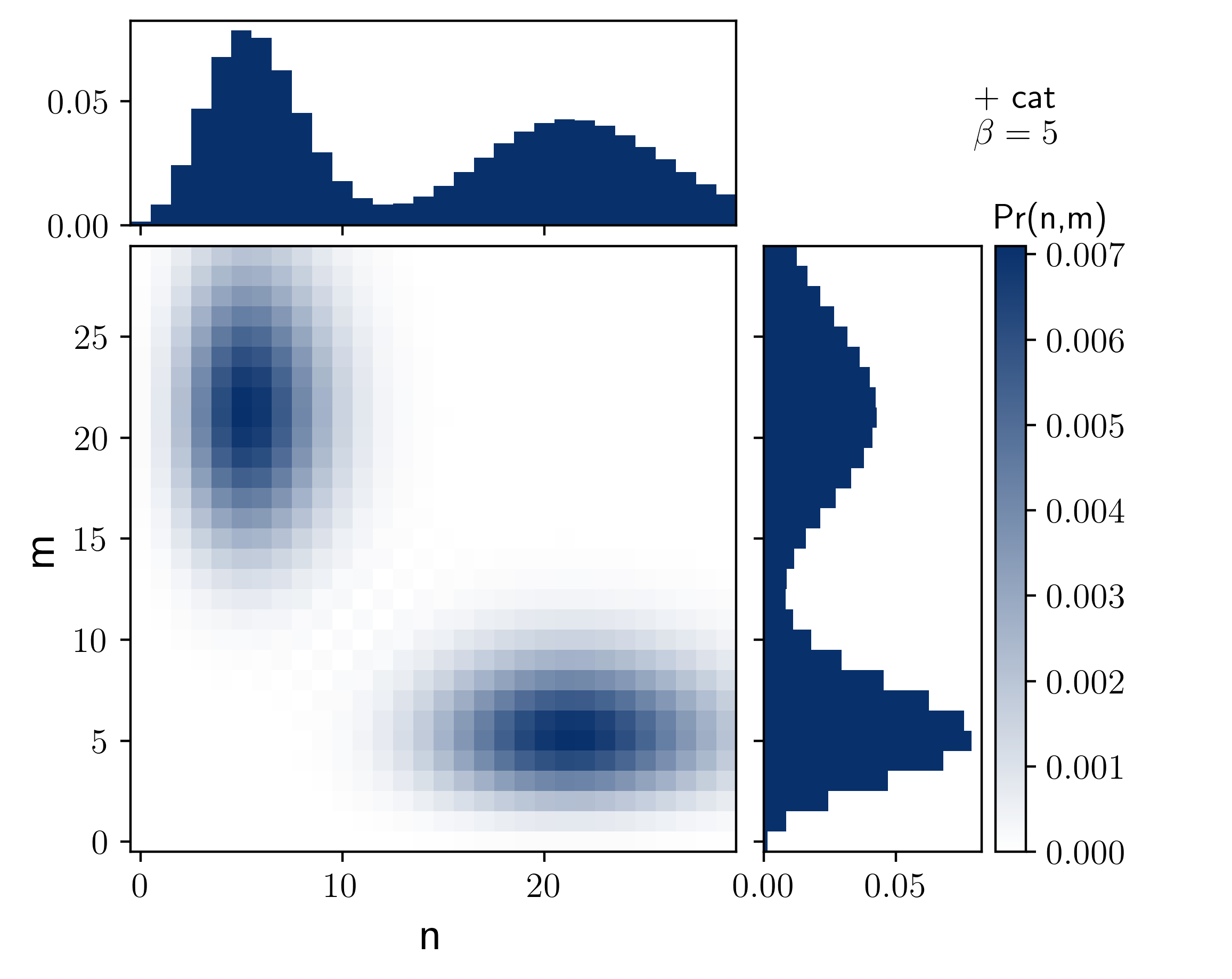} \\
    \includegraphics[width=\cohfiguresize]{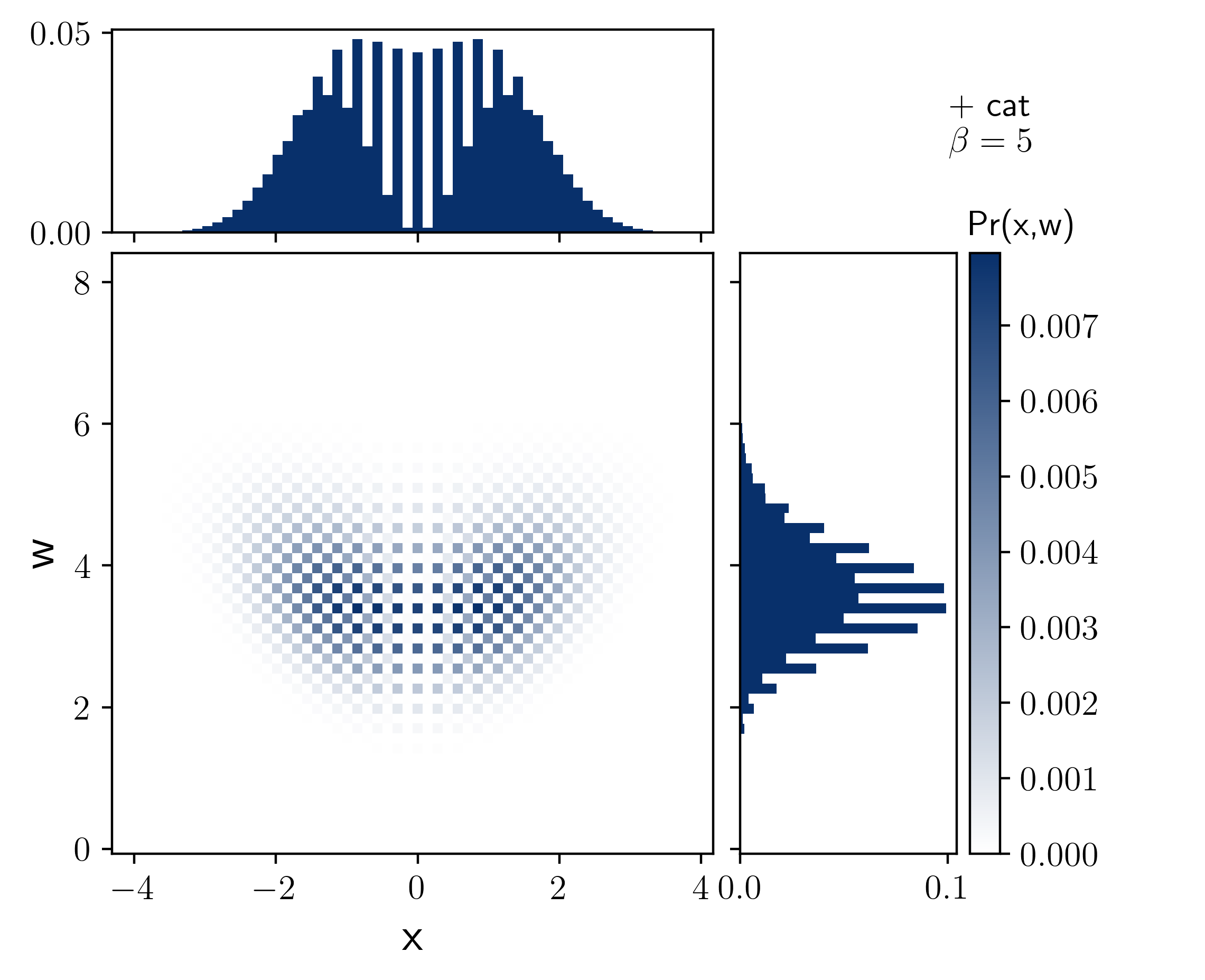}
    \includegraphics[width=\cohfiguresize]{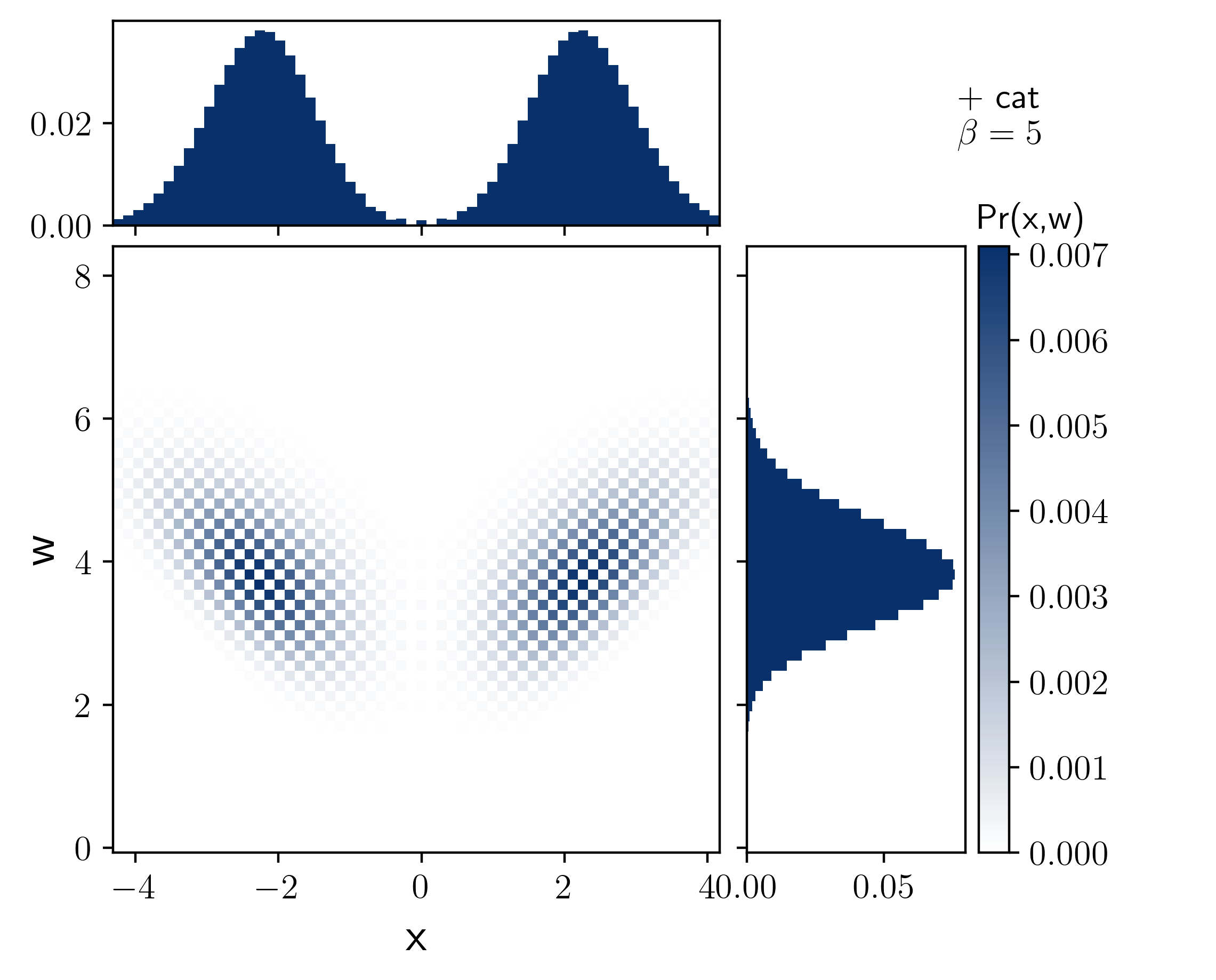}
    \includegraphics[width=\cohfiguresize]{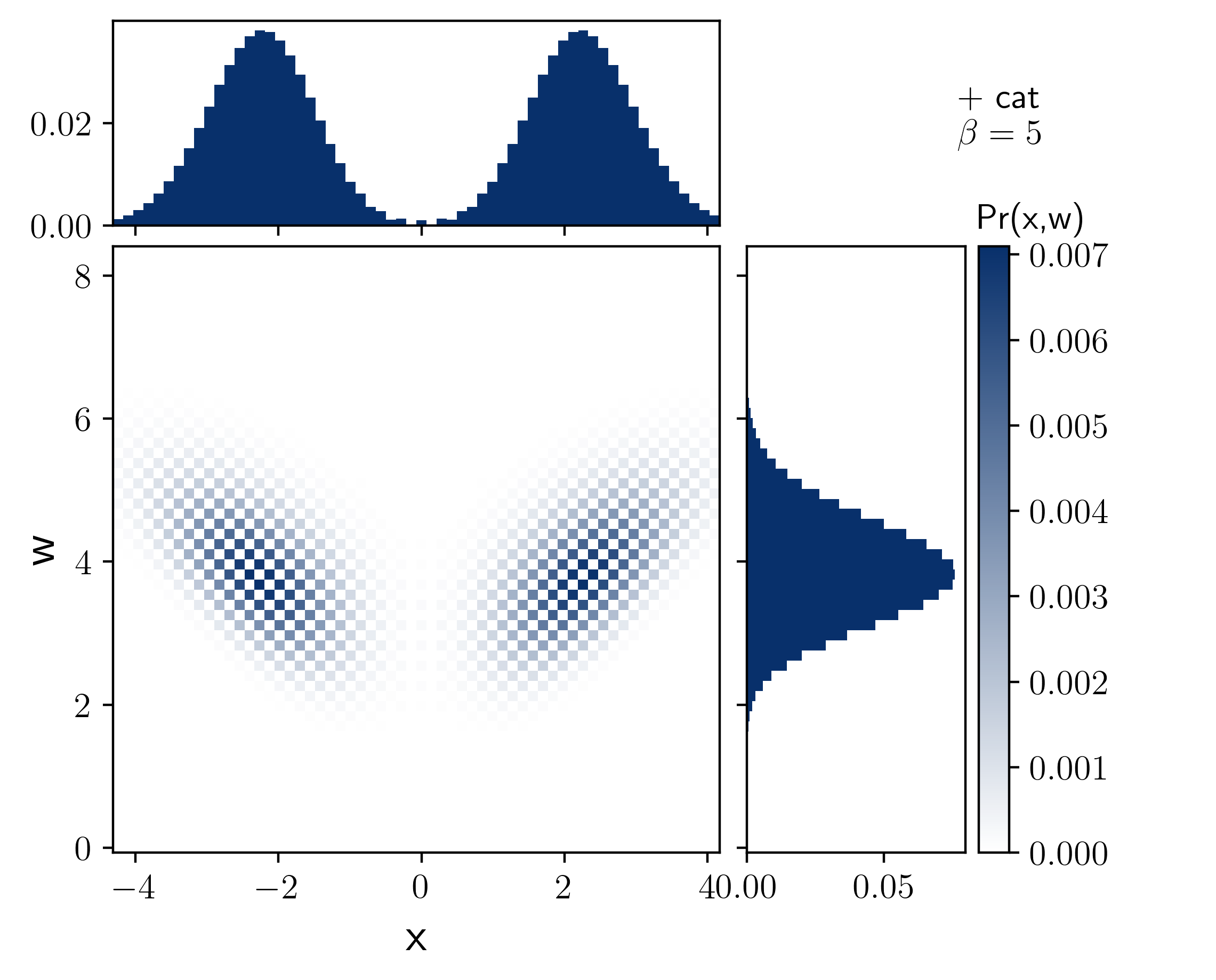} \\
    \caption{Click distributions for detecting a coherent state signal with  $\alpha=0.8$ (column 1), $\alpha = 1.6$ (column 2), and $\alpha = -1.6$ (column 3) with a Cat local oscillators where $\beta = 5$ and $\expt{n_b} = 25$. (Row 1) Original click distributions and marginal distributions for $n$ and $m$ {(dimensionless)}. (Row 2) Sum $w=(n+m)/(\sqrt{2}\beta)$ and difference $x=(n-m)/(\sqrt{2}\beta)$ variables {(dimensionless)} and corresponding marginal distributions.  In contrast to the previous plot the difference variables now contain little probability around $x=0$. Column 2 and 3 are identical as this measurement can not distinguish between $\ket{\alpha}$ and $\ket{-\alpha}$.}
    \label{fig:coherent_state_plots}
\end{figure*}

\subsubsection{Vacuum}

For the case of a vacuum input state as the signal, the exact Kraus operator from \cref{eq:exactcat_Kraus} acting on a vacuum state gives
\begin{multline}
    M_{n,m}^{[\beta]_\pm} \ket{0} = \\
    \frac{ e^{-|\beta|^2/2} } {2^{(n+m)/2} \sqrt{n!m!} \N_\pm(\beta)}
    \beta^{n+m} \left[(-1)^m \pm (-1)^n \right],
\end{multline}
where $(1\pm a/\beta) \ket{0} = \ket{0}$ and $(1-a^2/\beta^2) \ket{0} = \ket{0}$ have been used to form this expression.  The probability of detecting $n$ and $m$ photons is subsequently
\begin{multline}
\label{eq:vacuumprob}
    \Pr(n,m |\beta,\pm,\ket{0}) = \\
    \frac{e^{-|\beta|^2/2}}{n!} \left(\frac{|\beta|^2}{2}\right)^n \frac{e^{-|\beta|^2/2}}{m!}\left(\frac{|\beta|^2}{2}\right)^m  \\
    \left[\frac{(-1)^{m-n}\pm 1}{\N_\pm(\beta)} \right]^2.
\end{multline}
This equation shows explicitly the underlying Possion distribution envelops with equal Possion variables in both detectors. The signal will be modulated by the alternating interference in the final term in the square brackets.   This term comes entirely from the superposition state of the local oscillator.

Now one can change variables into the $x$ and $w$ sum and difference variables for this particular case.  However, at this stage, as the continuum approximation has not been made, the expression for this is not much clearer.  However, it should be noted that the final term in the square brackets \emph{only} depends on the difference variable $x$.

In \cref{fig:vacuum_plots} we plot the probabilities generated by both the ``$+$'' cat (i.e. $\ket{{\rm LO}}\propto \ket{\beta} + \ket{-\beta}$) and the ``$-$'' cat (i.e. $\ket{{\rm LO}}\propto \ket{\beta} - \ket{-\beta}$) as a LO when measuring the vacuum. The in the original photodetection variables $n,m$ (top row) the clicks of each variable individually are approximately Possion distributed with half to LO intensity $|\beta|^2/2$ due to the 50:50 beamsplitter.  The effects from the superposition of the local oscillator are present in the full distribution of $n$ and $m$ where complete interference gives zero probability from the final term in \cref{eq:vacuumprob}.  This gives rise to the ``checker-board'' style pattern in the full distribution and the exact terms where the pattern is non-zero depending on the phase of the LO cat-state superposition.  

In the sum and difference variables $x,w$ (bottom row) the two marginal distributions exhibit the interference more directly as they essentially look "diagonally" across the $n,m$ distribution.  The superposition in the local oscillator is evident in these distributions having non-zero probabilities only on odd or even number term depending on the sign of the local oscillator superposition.  Note that when changing to $x,w$ variables there exist particular combinations of variables that are permitted for individual $x$ and $w$ but not necessarily when combined together.  For example, if $x=0$ then $m=n$ and hence $m+n$ must be an even number and $w$ is an even multiple of $1/(\sqrt{2}\beta)$. But if $x=1/(\sqrt{2}\beta)$ (e.g. if $n=1,m=0$ or $n=3,m=2$, etc), then $n=m+1$ and $n-m$ is odd.  Hence $x=1/(\sqrt{2}\beta)$ with $w=1/(\sqrt{2}\beta)$ is not permitted.  As, depending on the superposition sign in the LO, particular parities of photon number are suppressed, this leads to stripes in the $x$ and $w$ marginal distributions.  These stripes are shifted by $1/(\sqrt{2}\beta)$ between the $+$ and $-$ superpositions in the LO.  This shifting occurs as the vacuum state has a definite even parity.  Therefore the parity of the sum and difference variables needs to preserve the overall parity relationship between the combined signal and LO.

After the information of the sum variable $w$ is integrated out, the stripes still remain.  This is because the parity of the sum and difference variables is determined by the parity of the input states only.  This is unlike the measurement in the raw counts $m$ and $n$ where the parity is encoded between the measurement outcomes as well as the parity of the input state.  Therefore the transformation to the sum and difference variables gives measurement outcomes that relate the measurements of parity even after integrating out one of the variables.
However, the parity information is encoded in outputs that are separated by $1/\sqrt{2}|\beta|$ which is equivalent to single integer changes in the $n$ or $m$ variables.  Any process that influences these numbers by a single integer, such as photon loss, would drastically reduce the visibility of this property.

This property of similar marginal distributions shifted by $1/(\sqrt{2}\beta)$ is commonly shared with different possible input signals.  For this reason, for other input states, we will only plot the $+$ superposition case of the local oscillator.  With the input state being the vacuum, the marginal distributions between $n$ and $m$ or between $x$ and $w$ look very similar.  This situation will change as we look a measuring signals from coherent light.

\subsubsection{Coherent state}

For the case of a coherent state input to the detector with the superposition local oscillator
the amplitude generated from the Kraus operator is
\begin{multline}
    M_{n,m}^{[\beta]_\pm} \ket{\alpha} = 
    \frac{ e^{-|\beta|^2/2} e^{-|\alpha|^2/2} } {2^{(n+m)/2} \sqrt{n!m!} \N_\pm(\beta)} \\
    \left[(\alpha+\beta)^{n}(\alpha-\beta)^{m} \pm (\alpha-\beta)^{n}(\alpha+\beta)^{m} \right].
\end{multline}
The probability of detecting $n$ and $m$ photons is subsequently
\begin{multline}
    P_{n,m}^{[\beta]_\pm} (\ket{\alpha}) = 
     \frac{ e^{-|\beta|^2-|\alpha|^2} } {n!m! \N_\pm(\beta)^2} \\ 
      \left|\left(\frac{\alpha+\beta}{\sqrt{2}}\right)^{n}\left(\frac{\alpha-\beta}{\sqrt{2}}\right)^{m} \pm \left(\frac{\alpha-\beta}{\sqrt{2}}\right)^{n}\left(\frac{\alpha+\beta}{\sqrt{2}}\right)^{m} \right|^2.
\end{multline}
These probabilities are shown in \cref{fig:coherent_state_plots} for input signal amplitudes of $\alpha = 0.8, 1.6$ and $-1.6$ and coherent state superpositions for the local oscillators with $\beta=5$ just like in \cref{fig:vacuum_plots}.

One of the striking things to notice in \cref{fig:coherent_state_plots} are the spikes in the sum and difference variables of the $\alpha =0.8$ signal, which might be an artefact of using photon number resolving detectors to approximate an intensity measurement.
Although, similar spikes are present in the work of \citet{SandLeeKim95} which considered a coherent state LO interfering with a cat state signal and in the marginal distribution of a cat state Wigner function. For larger intensity signals  e.g. $\alpha = 1.6$ we see the marginal distributions have smoothed out significantly. Importantly in columns 2 and 3 we can see that our measurement does not allow one to distinguish between $\ket{\alpha}$ and $\ket{-\alpha}$. In the large LO limit the (marginal) distribution of the $x$ variable seems limit to
\begin{equation}
    \Pr(x|\alpha) = \Tr [{E_x  \op{\alpha}{\alpha}}] = \frac 1 2 \big ( |\ip{\alpha}{x_\theta}|^2  +  |\ip{\alpha}{-x_\theta}|^2 \big ),
\end{equation}
where $\ip{\alpha}{x_\theta}$ is the inner product between a coherent state and a rotated quadrature eigenstate. This expression also holds when $\alpha =0$ as in the previous section. Of course this is not the case for the first column where the interference terms are still visible.

In \cref{apx:LO_size} we study the effect of small LOs, relative to the signal strength, on a coherent state signal. We find numerically that LO's slightly larger than the signal might be sufficient to enable our proposed measurement.

\newcommand{\fockfiguresize}{0.67\columnwidth}
\begin{figure*}[htb]
    \centering
    \includegraphics[width=\fockfiguresize]{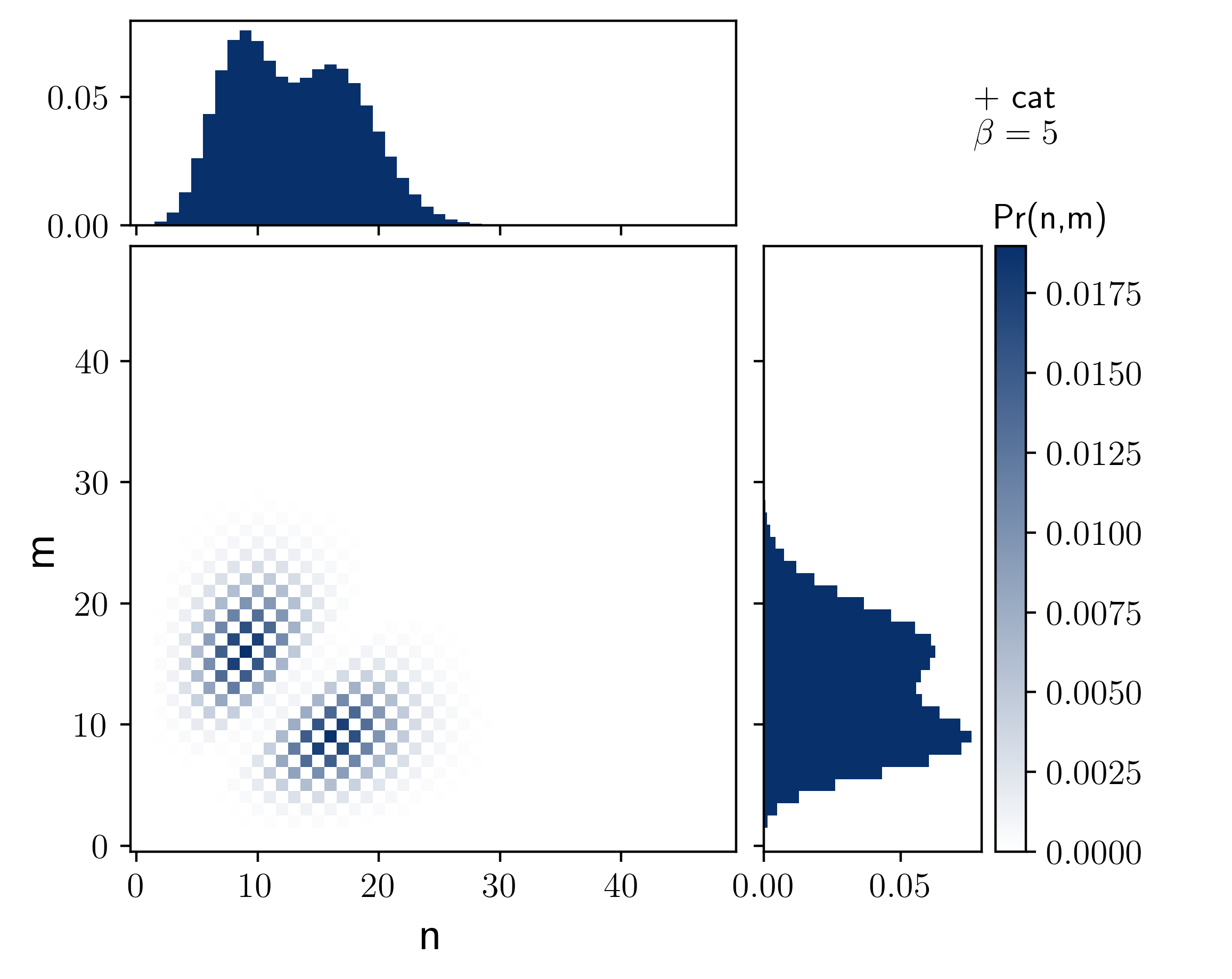}
    \includegraphics[width=\fockfiguresize]{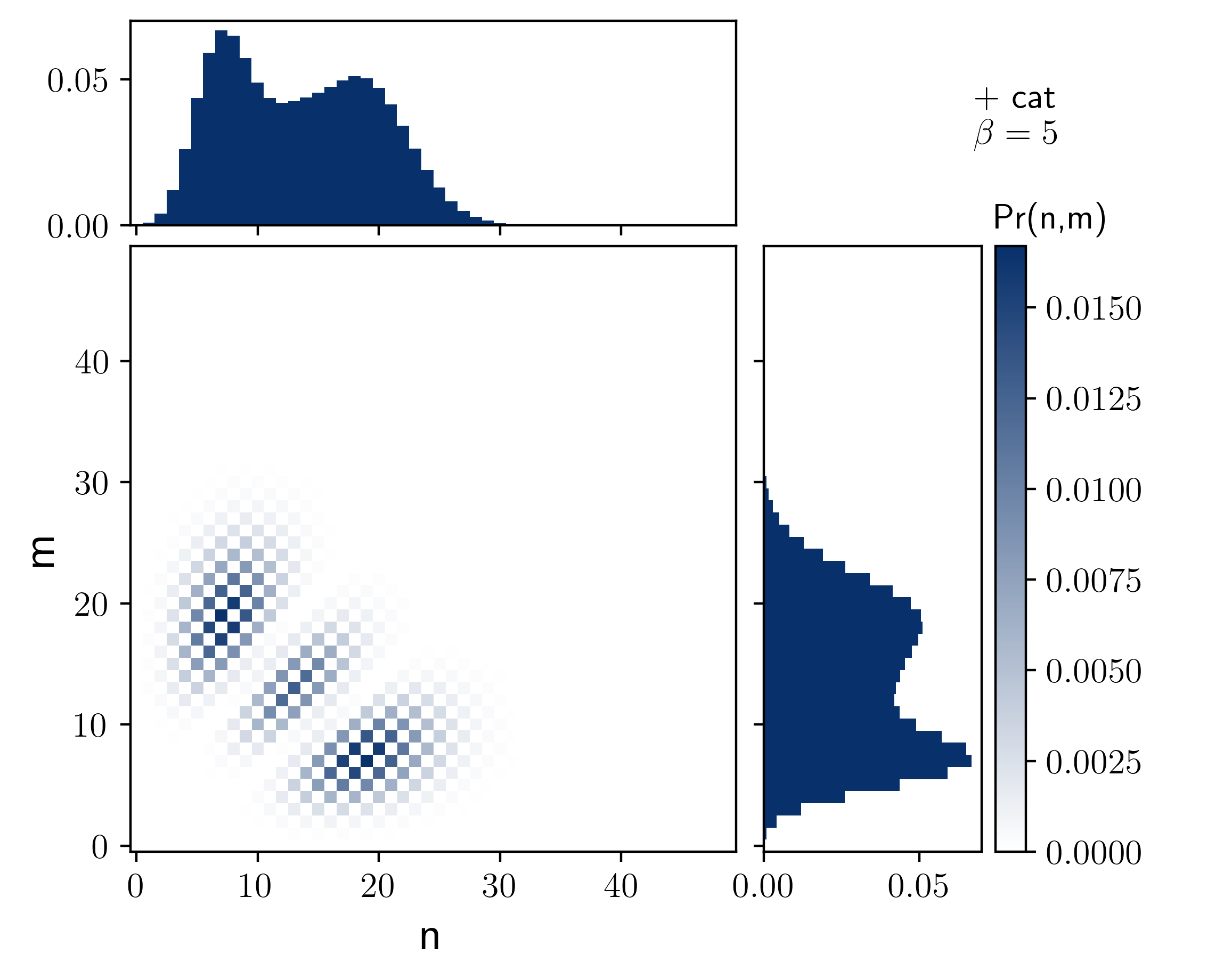}
    \includegraphics[width=\fockfiguresize]{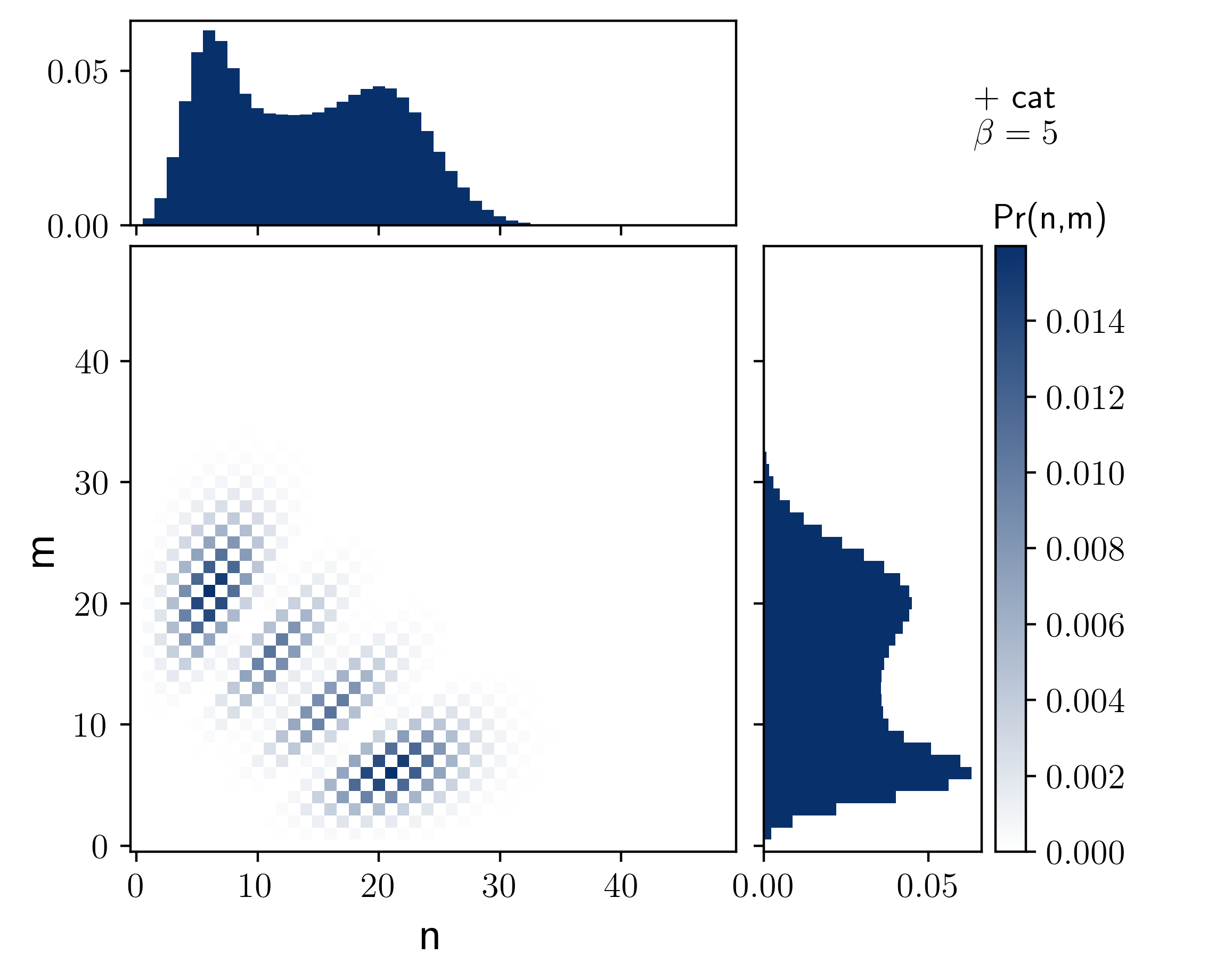}\\
    \includegraphics[width=\fockfiguresize]{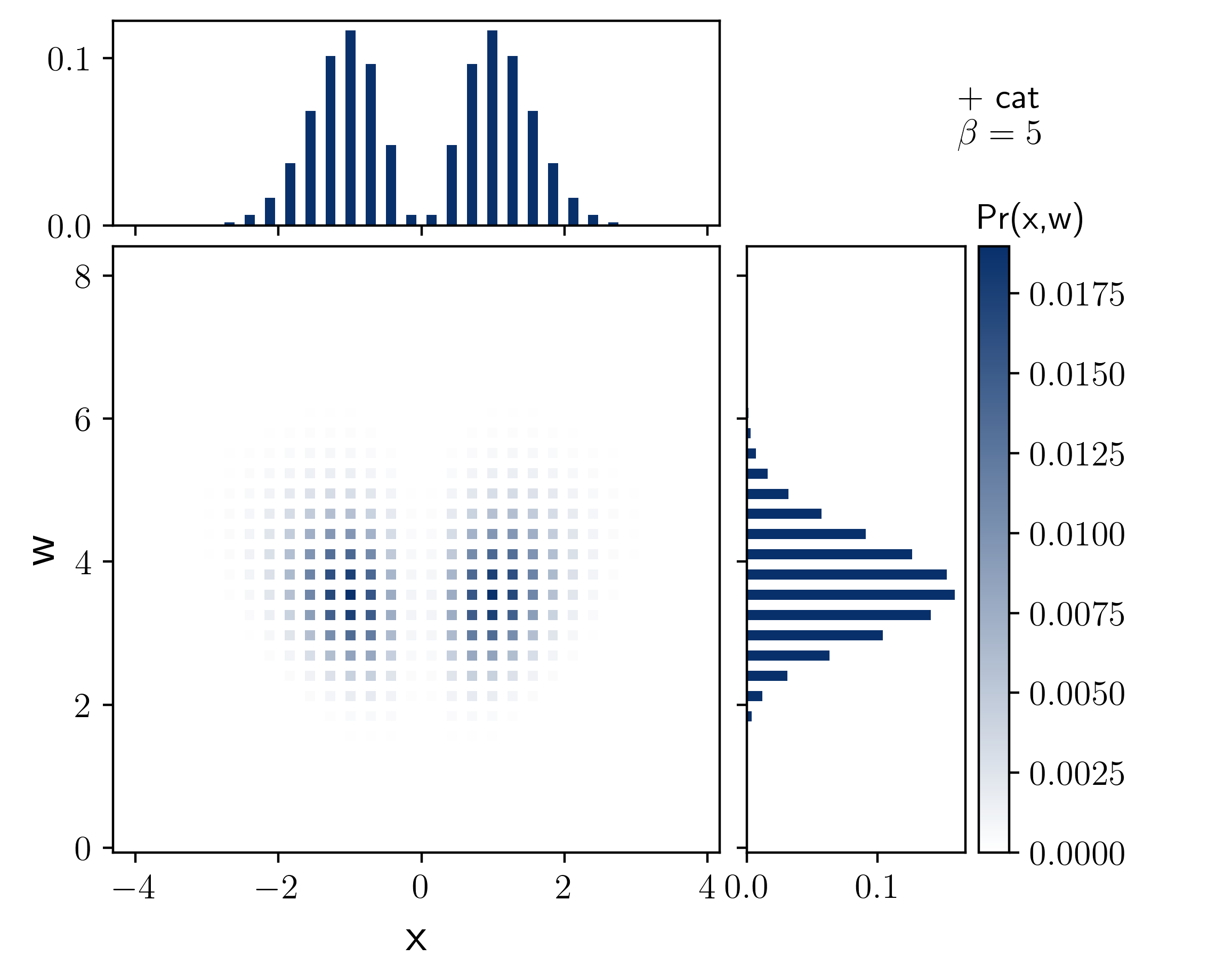}
    \includegraphics[width=\fockfiguresize]{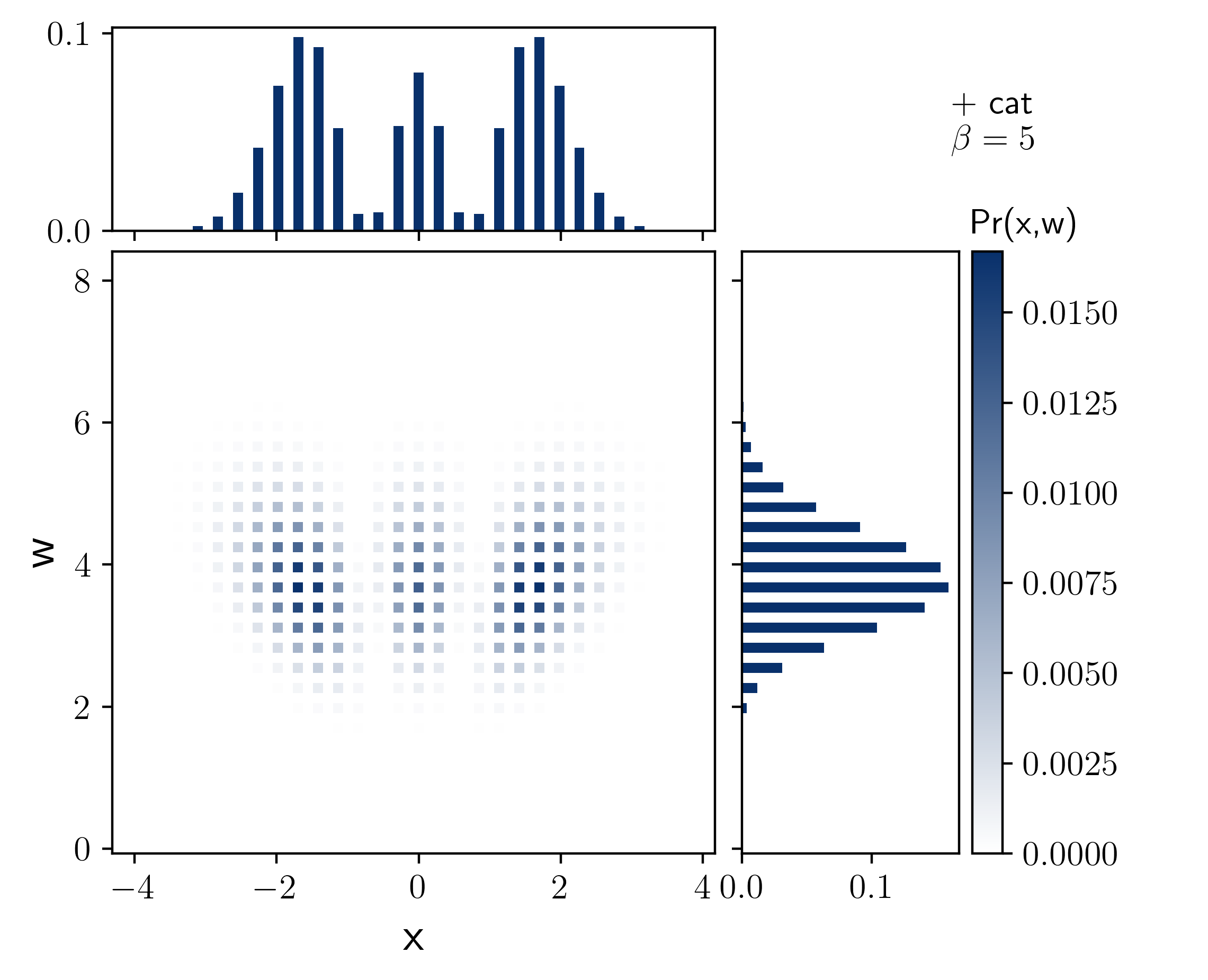}
    \includegraphics[width=\fockfiguresize]{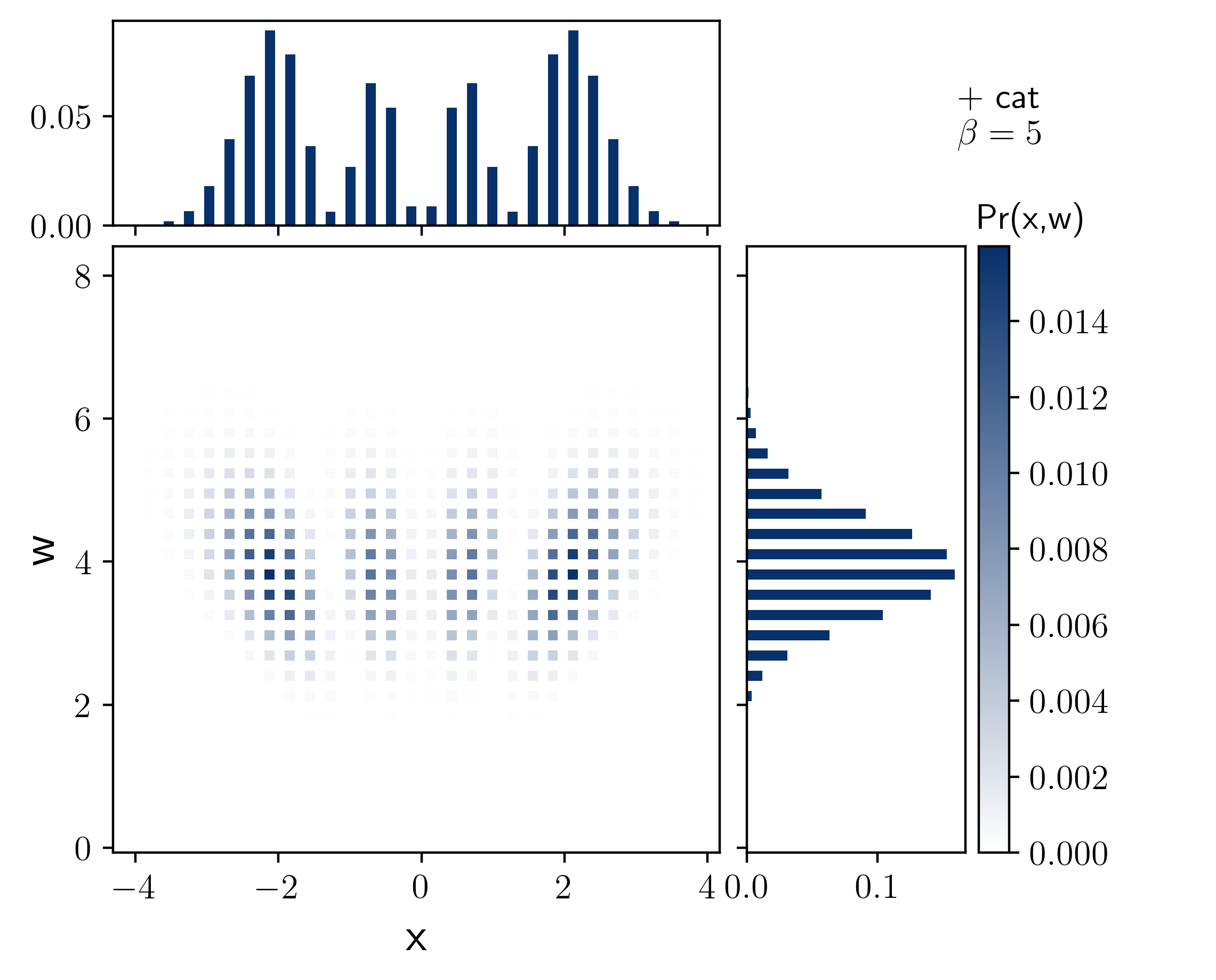}
    \caption{Click distributions for detecting a Fock state $\ket{\focksym}$ with  $\focksym=1$ (column 1), $\focksym = 2$ (column 2), and $\focksym = 3$ (column 3) with a Cat local oscillators where $\beta = 5$. (Row 1) Original click distributions and marginal distributions for $n$ and $m$ {(dimensionless)}. (Row 2) Sum $w=(n+m)/(\sqrt{2}\beta)$ and difference $x=(n-m)/(\sqrt{2}\beta)$ variables {(dimensionless)} and corresponding marginal distributions.}\label{fig:fock_plots}
\end{figure*}

\subsubsection{Fock state}

For the case of a Fock state input state, i.e. $n \ket{\focksym} = \focksym \ket{\focksym}$, as the signal, equation~\ref{eq:exactcat_Kraus} acting on such a state gives
\begin{multline}\label{eq:fockfragment}
    M_{n,m}^{[\beta]_\pm} \ket{\focksym} =
    \frac{ e^{-|\beta|^2/2} } {2^{(n+m)/2} \sqrt{n!m!} \N_\pm(\beta)} \\
    \bra{0} \left( (\hat{a}+\beta)^n (\hat{a}-\beta)^m \pm (\hat{a}-\beta)^n (\hat{a}+\beta)^m\right)\ket{\focksym}.
\end{multline}
In \cref{apx:fock} we show how to simplify this expression. the resulting closed form helps mainly with numerical computation but offers little insight into the general functional properties. In the large LO limit the distribution of the $x$ variable should limit to
\begin{equation}
    \Pr(x|\focksym) = \Tr [{E_x  \op{\focksym}{\focksym}}] = \frac 1 2 \big ( |\ip{\focksym}{x_\theta}|^2  +  |\ip{\focksym}{-x_\theta}|^2 \big ),
\end{equation}
where $\ip{\focksym}{x_\theta}$ is the inner product between a Fock state and a rotated quadrature eigenstate. This in itself will be proportional to the square of a Hermite polynomial $H_\focksym(x)$ as
\begin{equation}
\ip{x}{\focksym} = \frac{1}{\sqrt{2^\focksym \focksym!}} H_\focksym(x) \ip{x}{0}.
\end{equation}
In \cref{fig:fock_plots} the appearance of the square of a Hermite polynomial is evident in the difference variable marginal distributions. Also evident is the effect of parity on the distribution of the difference variable $x$. Columns 1 and 3 have odd parity input states ($\ket{1}, \ket{3}$), while column 2 has and even input parity ($\ket{2}$). Like in the vacuum case the even parity states have support on the difference variable when $x=0$ and the odd parity states do not.

% ===========================================================
\section{Application of catodyne to remote state preparation}\label{sec:remotecatprep}
% ===========================================================
We now briefly describe an application of our measurement to remotely preparing superposition of position eigenstates or a ``\sch\ Cat state in position''. The idea is to have two parties share an EPR state (which is a Gaussian state) and then perform our non-Gaussian ``catodyne'' measurement, i.e. \cref{eq:cat_kraus_almostdone}, on half of the EPR pair. Then, conditional on the measurement result $q,w$, the state
\begin{equation}\label{eq:peparedstate}
    \ket{c_{q,w}} { \propto} %\frac{1}{\sqrt{2}}(
    \ket{q} + (-1)^{f(w)} \ket{-q}
    %)
\end{equation}
is prepared remotely. In \cref{eq:peparedstate} $\ket{q}$ is an eigenstate of the position operator $x$ i.e. $x\ket{q}= q\ket{q}$, and $f(w)$ is a linear function of the sum variable $w$. This procedure is summarised in the following circuit.\vspace{-6pt}
\begin{center}
\begin{equation}
\includegraphics[width=0.75\columnwidth]{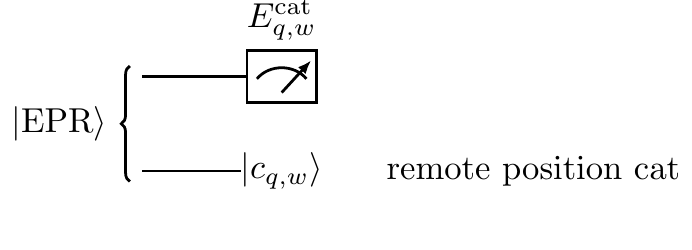}
\end{equation}
\end{center} \vspace{-4pt}
We will start by considering a simplistic case using an ideal EPR state (an infinitely squeezed, two-mode squeezed vacuum state) that illustrates the basic features of the protocol and then consider, a more realistic, finitely squeezed state. {The states resulting from the finite squeezing version are promising bosonic error correcting codes~\cite{schlegel_quantum_2022,xu_autonomous_2022}.} {We also note that this scheme is useful if the remote party only has access to linear detectors and linear passive elements such as beam splitters.}

%-------------------------------------------
\subsection{Infinite squeezing}\label{sec:epr}
The initial state between the remote parties is an ideal EPR state in the position representation
\begin{equation}\label{eq:EPR}
    \ket{\rm EPR} = \int dx \,  \ket{x}\otimes \ket{x},
\end{equation}
which is unnormalizable and unphysical; but a limit of states that are routinely made in the optical domain using two mode squeezing.

We will choose to measure one of the systems and allow the other to freely propagate. To compute the state of both systems after the measurement we use the usual measurement update rule
\begin{equation}\label{eq:measrule}
\ket{\Phi_{q,w}} = \frac{ (M_{q,w}^{[+]}\otimes \Id   ) \ket{\rm EPR} }
{\sqrt{\expt{{\rm EPR} |E_{q,w}^{[+]} | {\rm EPR}}}}.
\end{equation}
Since the denominator simply normalizes the post-measurement state it is instructive to consider the numerator of \cref{eq:measrule} alone
\begin{align}
(M_{q,w}^{[+]}\otimes \Id   ) \ket{\rm EPR} \propto (\bra{q} \pm \bra{-q}) \otimes \Id   \int dx  \ket{x}\otimes \ket{x} \, ,
\end{align}
where we have replaced the $w$ dependence of $M_{q,w}$ with a parity bit $\pm$ to simplify our presentation. Moreover it is clear that the measured mode is absorbed by the detector. Performing the integrals we arrive at
\begin{equation}\label{eq:unphys_cat}
\ket{\Phi_{q,w}} { \propto} %\frac{1}{\sqrt{2}}(
\ket{q} + (-1)^{f(w)} \ket{-q}\, ,
%),
\end{equation}
which is the \sch\ cat state in position on the remote system. Of course, position eigenstates are not normalizable but are a limit of single mode squeezed states that are normalizable.  In anycase, we rectify this simplistic treatment in the next section with normalizable states that limit to this result.

%-------------------------------------------
\subsection{Finite squeezing}\label{sec:finite}
In the position representation, a general pure two-mode state can be expressed as
\begin{equation} 
\ket{\psi} = \int dx_a \int dx_b\, \psi (x_a,x_b)\ket{x_a}\otimes \ket{x_b}
\end{equation}
where $\psi (x_a,x_b)$ is the two mode position wavefunction. 
A two-mode (finitely) squeezed vacuum state is given by the position wavefunction (see Eq. (81) in Ref.~\cite{Braunstein_vanLoock_RMP})
\begin{equation}\label{finiteSqeezing}
\psi_{r}(x_a,x_b)_{\rm TMSV}=\sqrt{\frac{2}{\pi}} e^{-e^{-2r}(x_a+x_b)^2/2}e^{-e^{2r}(x_a-x_b)^2/2},
\end{equation}
where $r$ is the squeezing factor.  In the limit $r\to\infty$, the wavefunction limits to $\psi_{r}(x_a,x_b) \propto \delta(x_a-x_b)$ which gives rise to the EPR state in \cref{eq:EPR}. 

\begin{figure}[htb]
\centering
    \includegraphics[width=0.99\columnwidth]{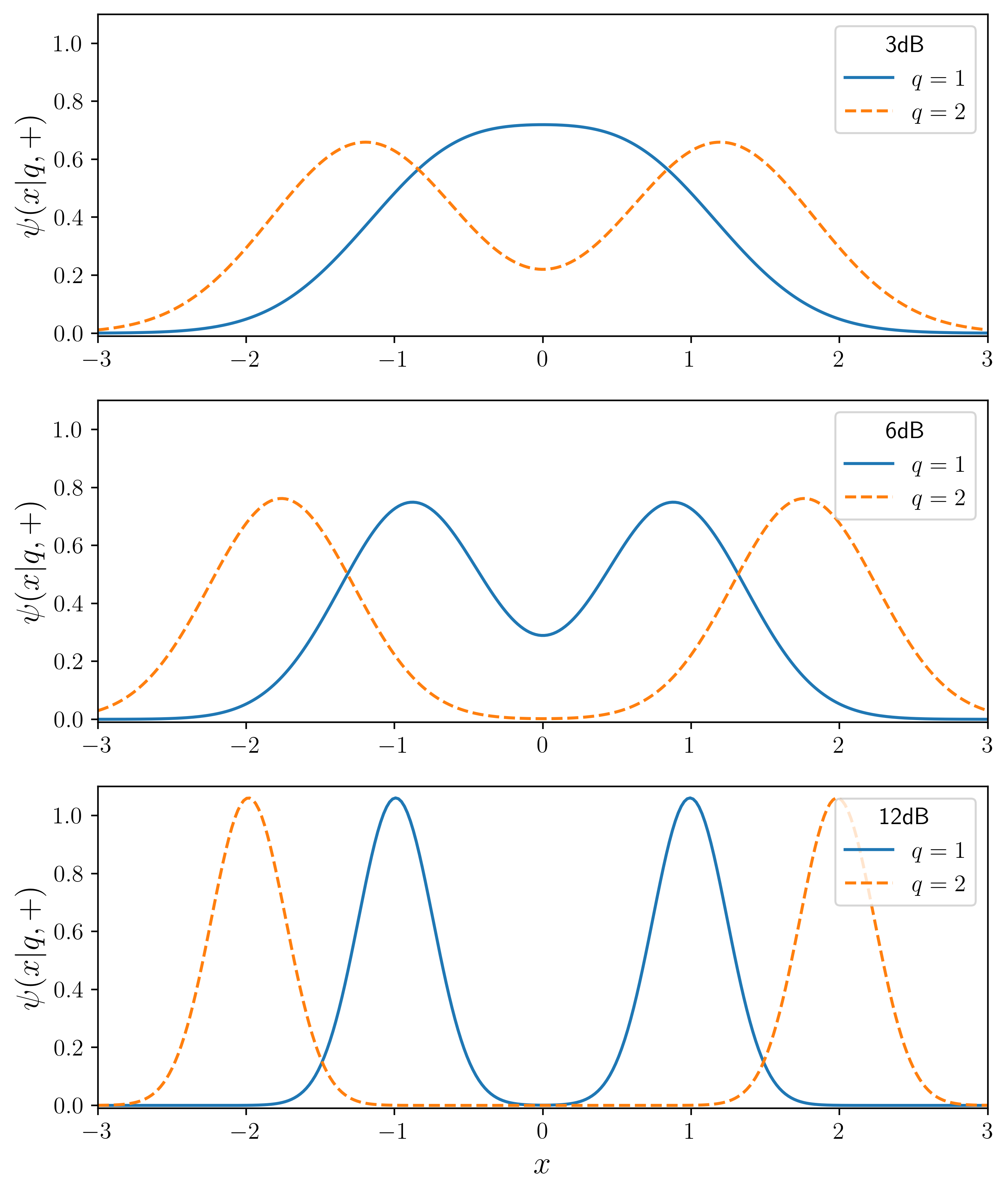} 
    \caption{ Position wave function $\psi(x|q,+)$, i.e. \cref{eq:postmeas_finite_squeeze}, of remote state given measurement results $q = 1$ or $q=2$. (Top) At 3dB of squeezing, i.e. $ r\approx 0.345$ the remote state is barely a cat state in position for $q=2$ measurement result.  (Middle) 6dB of squeezing or $r\approx 0.691$. Now we can see the emergence of a superposition for both outcomes. (Bottom) 12dB of squeezing or $r\approx 1.382$. Now we have clear cat states for both outcomes and the Gaussian distributions are centred on $q=1$ or $2$. Indeed as $r\rightarrow \infty$ we limit to \cref{eq:unphys_cat}. Note: {we are using dimensionless versions of the quadrature operators so  $x$ is a dimensionless} and  to convert to squeezing in dB we use using $r_{\rm dB} =10 \log_{10}e^{2r}$.  }\label{fig:remote_wfn}
\end{figure}

As before we consider an ideal ``catodyne'' measurement, i.e. the Kraus operators in \cref{eq:cat_kraus_almostdone}, on mode ``a'' which leaves a conditional state on mode ``b''. With some simplification, detailed in \cref{apx:EPR_twomode_squeeze}, the normalized post measurement position wavefunction after obtaining measurement outcome $\{q,\pm\}$ is
\begin{align}\label{eq:postmeas_finite_squeeze}
\psi(x|q,\pm  & ) = \left (\frac{\cosh 2 r}{2 \pi }\right ) ^{1/4} 
\frac{e^{q^2 \sinh 2 r \tanh 2r }}{\sqrt{e^{2 q^2 \sinh 2 r \tanh 2 r}} \pm 1} \times \\
&\left [e^{-\cosh 2r (x + q \tanh 2r)^2} \pm e^{-\cosh 2r (x  - q \tanh 2r)^2 } \right ], \nonumber
\end{align} 
where $q$ is the position outcome, $\pm$ is the plus or minus parity outcome, and $r$ is the squeezing parameter from \cref{finiteSqeezing}. This squeezed cat state limits to  \cref{eq:unphys_cat} as $r\rightarrow \infty$ because both terms inside the square brackets are Gaussian.
To get an intuitive idea of interplay between the amount of squeezing $r$ and $q$ measurement outcome we plot $\psi(x|q,+)$ in  \cref{fig:remote_wfn} 
 for two measurement outcomes $q=1$ and $q=2$ for three levels of squeezing.

\begin{figure}[htb]
\centering
    \includegraphics[width=0.99\columnwidth]{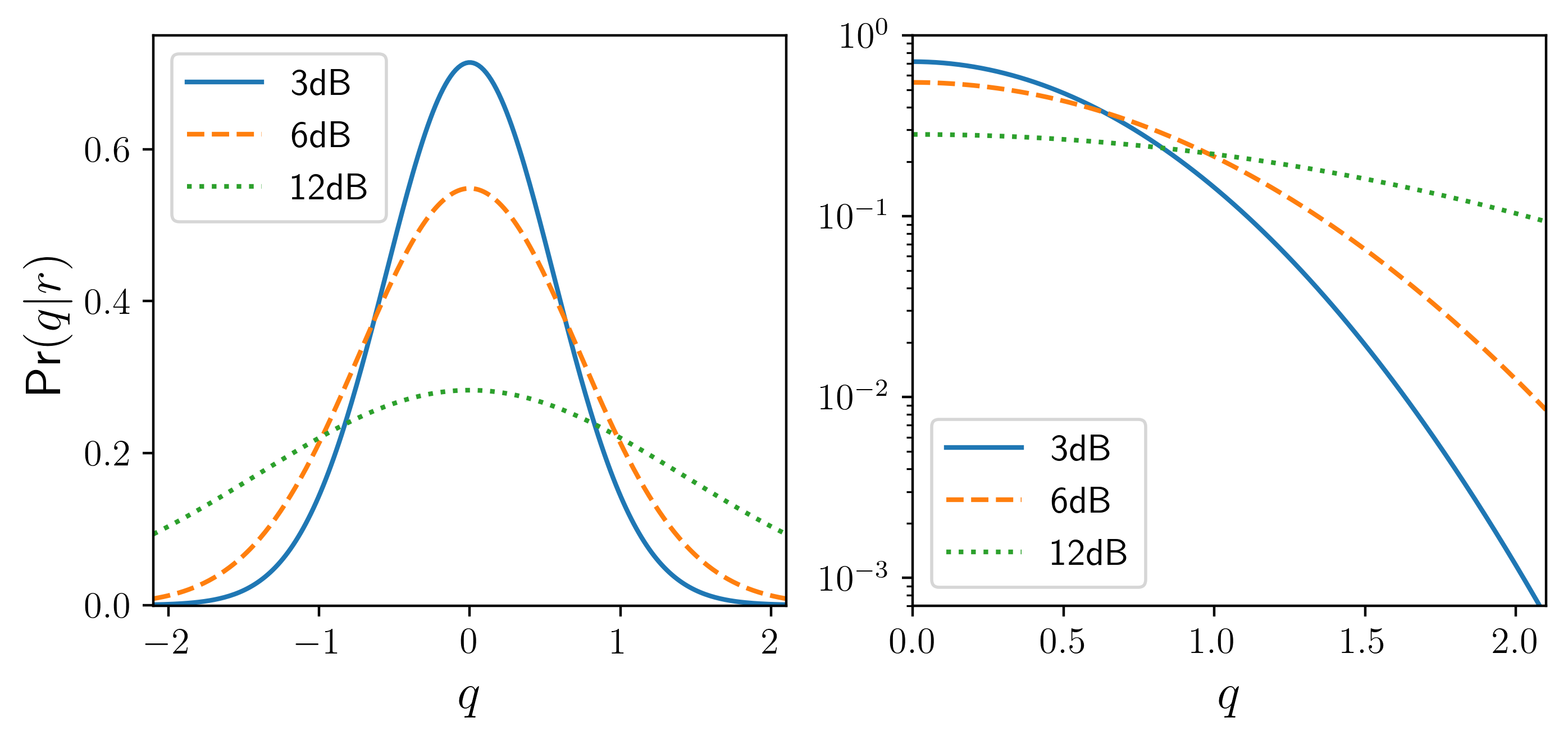}
    \caption{ The probability $\Pr(q|r)$ of obtaining a measurement outcome $q$ for different amounts of squeezing i.e.  \cref{eq:pr_q_finite_squeeze}. (Left) a linear plot of $\Pr(q|r)$. (Right) a linear-log plot of $\Pr(q|r)$ for $q\ge 0$. Evidently, the probability for post selecting on $q =2 $ for 3dB of squeezing is around $10^{-3}$, making this an unlikely event. At 12dB of squeezing many values of $q$ have a reasonable chance of occurring. }\label{fig:prq_squee}
\end{figure}

The probability of obtaining the outcome $q$ is normally distributed in $q$ around $q=0$
\begin{equation}\label{eq:pr_q_finite_squeeze}
dq\,\Pr(q|r)= dq \sqrt{\frac{2 \sech 2 r}{\pi }}e^{-2 q^2 \sech 2 r}.
\end{equation} 
As the variance is proportional to $\sech 2r = 1/\cosh 2r$ it is clear that larger squeezing will stop penalising large $q$ values.
In \cref{fig:prq_squee} we plot $\Pr(q|r)$ for several values of $r$.

The main take-home message from \Cref{fig:remote_wfn,fig:prq_squee} is that more than 3dB of squeezing is likely needed for a robust  demonstration a remote state preparation protocol using our measurement. This shouldn't be a problem as modern experiments have demonstrated squeezing in the 15dB range~\cite{15dB_squeezing_experiment}.

Finally we should point out that our analysis did not take into account a finite strength LO or imperfections in number resolved detection, which is a good topic for future work to study. Nevertheless preliminary results, see \cref{apx:LO_size}, indicate that LOs that are moderately larger than signal may be adequate for realising our scheme. The resulting POVM would not be a projection onto superpositions of quadrature eigenstates but (likely) a projection onto superpositions squeezed states. 

% ===========================================================
\section{Conclusion}\label{sec:conc}
% ===========================================================
Using a non-classical local oscillator we have constructed a new non-Gaussian measurement. Two new features that arose, relative to standard homodyne measurements, were the importance of number resolved detection and the importance of information contained in the sum variable. Using the {\em number resolved} information from the sum variable led to a measurement that had coherence between outcomes (a rank-1 POVM), while integrating out the sum variable led to a loss of coherence (a rank-2 POVM) which could be achieved by a mixed state local oscillator.

We have shown that these measurements can be used to remotely prepare a non-Gaussian state. While we did not get non-Gaussiannity for free, we injected a LO that was non-Gaussian and the number resolving measurements are non-Gaussian. Nevertheless, the ability to prepare non-Gaussian states via measurement in a teleportation scheme might find applications in quantum computation and communications. The utility of these measurements outside it's usefulness in remote state preparation is unknown. However it is heartening to note that ordinary homodyne (or heterodyne) measurements can be used to measure nonlinear properties such as correlation functions \cite{SilvaBozy2010}. Thus it is possible that the measurements described herein may find useful and exotic applications.

So far, we have not discussed the experimental feasibility of our scheme, let us separate the discussion into the microwave and optical domains. 
% microwave
In the microwave domain, it seems likely that our scheme could be realised now as generation non-classical states~\cite{Krastanov2015,Heeres2017} and photon number detection~\cite{Schuster2007} inside a cavity are routine. One realisation would be to  prepare the signal state in one microwave cavity and the LO in another and interfere them (via a beam splitter interaction) and then do number resolved detection on both. A microwave frequency travelling wave demonstration of our protocol would be significantly harder.
% optical
In the optical domain, small cat states have been experimentally demonstrated \cite{Gerrits2010,Serikawa2018} and protocols for turning small cats in to larger cats (aka ``breeding'') have been explored for a number of years~\cite{Lund2004,Laghaout2013,Sychev2017} and photon number resolving detectors can resolve more than 15 photons with reasonable efficiency~\cite{morais_precisely_2022}. Moreover, having small cat states may be desirable, and not a limitation, as alluded to in the main text. Nevertheless demonstrating all components together appears to be technologically challenging at present.

There are many possible extensions of this work for example one could consider other non-classical states where you can take a large LO limit. 
Moreover, we have not numerically or analytically studied the convergence of the full measurement operators to the strong local oscillator limit~\cite{Braunstein90,Tyc_2004}. An important open question is how large does the local oscillator have to be a reasonable approximation to the limiting measurement operators. A second and equally important question is the extension of our single mode analysis to the inefficient detection of multimode fields ~\cite{Shapiro85,Collett87,Barchielli_1990,BraunCrou91, OuKimble95, SasakiSuzuki06} which would enable the study of mode matching effects. Some stepping stones to this end have been made by e.g. \citet{Gough2012} and \citet{Dabrowska2019} where a quantum trajectory formulation of fields in superpostions of coherent states was derived. These effects will play a central role in any experimental realization of our scheme.

Some of the artefacts we saw in the numerics were due to the fact we considered an idealized situation where we had number-resolving detectors. In practice, Homodyne detection is achieved by intensity detectors. An approximation of an intensity measurement would be to use the finite-efficiency photon number POVM. Given an efficiency $\eta\in [0,1]$ the measurement operator representing report $n$ clicks is
\begin{equation}
    E_n^{(\eta)}\! =\! \sum_{m=0}^\infty \binom{n+m}{n} \eta^n (1-\eta)^m \op{n+m}{n+m}.
\end{equation}
Using this operator in the analysis should remove the spurious issues with parity of the $w$ variable we noted in \cref{sec:cat} and be closer in spirit to true Homodyne detection. However it is unclear if this will result in a rank 1 or rank 2 POVM. \\

\noindent {\em Acknowledgements}: The authors thank Ben Baragiola, Anita D{\k a}browska, Mac Kim, Noah Lordi, Nicolas Menicucci, and Eugene Tsao for helpful discussions. We would also like to thank both anonymous referees for helping us to clarify and improve the manuscript.
% Josh
JC was funded in part by the Australian Research Council, through the Discovery Early Career Research Award (DECRA) project number DE160100356, the National Science Foundation QLCI Award No. OMA--2016244,  and Office of Naval Research award No. N00014-22-1-2438. 
% Austin
APL acknowledges support from BMBF (QPIC) and the Einstein Research Unit on Quantum Devices. This research was supported by the Australian Research Council (ARC) under the Centre of Excellence for Quantum Computation and Communication Technology (Project No. CE170100012).

\begin{widetext}
% ============================================================
\appendix
% ============================================================

%-------------------------------------------------------------
\section{Coherent state signals for regular homodyne}\label{app:quad}
%-------------------------------------------------------------
In this section we use a coherent state, $\ket{\alpha}$, as a proxy for an 
arbitrary signal states with $\langle n_{\rm signal} \rangle \ll \langle n_{\rm LO} \rangle$ i.e. $|\alpha|^2\ll |\beta|^2$. This let's us reason about properties of the click distribution that are due, largely, to the LO.

If the input signal is an coherent state $\ket{\alpha}$ then
\begin{align}
	M_{n,m}^{[\beta]} \ket{\alpha} =  
	\frac{(\alpha+\beta)^n (\alpha-\beta)^m}
		{2^{(n+m)/2} \sqrt{n!m!}}
		e^{-(|\beta|^2 + |\alpha|^2)/2}
		\ket{0} 
		= 
		\frac{(\alpha+\beta)^n}{2^{n/2}\sqrt{n!}}
		e^{-(|\alpha+\beta|)^2/2} 
		\frac{(\alpha-\beta)^m}{2^{m/2}\sqrt{m!}}
		e^{-(|\alpha-\beta|)^2/2} 
		\ket{0}
\end{align}
which gives probabilities proportional to separate Poisson distributions
\begin{equation}
	P^{[\beta]}(n,m)=
	\frac{|\alpha+\beta|^{2n}}{2^{n}n!} e^{-(|\alpha+\beta|)^2} 
	\frac{|\alpha-\beta|^{2m}}{2^{m}m!} e^{-(|\alpha-\beta|)^2}.
\end{equation}
In the case of a strong local oscillator relative to the input signal, i.e. $\beta \gg \alpha$, then the distributions of $n$ and $m$ are individually peaked nearby a mean value of $|\beta|^2/2$ with standard deviation $\beta/\sqrt{2}$.

The mean difference between $n$ and $m$ (normalised by $\sqrt{2}|\beta|$) can be computed using the statistical moments of the Poisson distribution 
\begin{equation}
	\frac{{\rm E} ( n - m )}{\sqrt{2}|\beta|} 
         =
	\frac{|\alpha+\beta|^2}{2\sqrt{2}|\beta|} 
	-\frac{|\alpha-\beta|^2}{2\sqrt{2}|\beta|} 
	= \frac{e^{-i\theta} \alpha + e^{i\theta}\alpha^*}{\sqrt{2}}
\end{equation}
where ${\rm E} ( n - m )$ is the expectation of the difference variable and $\theta$ is the complex angle of $\beta = |\beta| e^{i \theta}$.  This recovers the quadrature component and shows how the signal output is converted into the quadrature signal.  The variance is then
\begin{equation}
	\frac{{\rm Var}(n - m)}{2|\beta|^2} = 
	\frac{|\alpha+\beta|^2}{4|\beta|} 
	+\frac{|\alpha-\beta|^2}{4|\beta|} 
	= \frac{|\alpha|^2 + |\beta|^2}{2|\beta|^2}
	= \frac{1}{2} + \frac{|\alpha|^2}{2|\beta|^2}
\end{equation}
as the variances add.  In the strong local oscillator limit the second term tends to zero and this variance approaches the variance of the input coherent state.

%-------------------------------------------------------------
\section{Error estimate of approximation}\label{app:error_est}
%-------------------------------------------------------------
We wish to expand a function $f$ of a random variable $X$ about the mean of $X$ i.e. ${\rm E}[X]$. The expansion is \cite{benaroya2005} 
\begin{align}
{\rm E}[f(X)] \approx f \Big ( {\rm E}[ X] \Big ) + \frac{1}{2} {\rm Var}[X] f''\Big({\rm E}[X]\Big)+ \ldots 
\end{align}
The function of the random variable $m$ we care about is
\begin{align}
f(m)=	\left(1-\frac{e^{-2i \theta}\hat{a}^2}{|\beta|^2}\right)^m \, .
%	\left(1-\frac{e^{-2i \theta}\hat{a}^2}{|\beta|^2}\right)^{|\beta|^2/2}.
\end{align}
Recall that $m$ is the number of clicks in one of the detectors. Because the LO overwhelms the signal, the number of clicks is approximately Poisson distributed and that means that ${\rm E}[m]= {\rm Var}[m]= \langle m \rangle = |\beta|^2/2 $. Further the second derivative of $(1 - x)^m$ with respect to $m$ is  
$(1 - x)^m \ln^2(1 - x)$. Combined this gives us

\begin{align}
{\rm E}[f(m)]\approx& 	\left(1-\frac{e^{-2i \theta}\hat{a}^2}{|\beta|^2}\right)^{|\beta|^2/2}  + 
\frac{1}{4} |\beta|^2 \left(1-\frac{a^2 e^{-2 i \theta }}{|\beta|^2}\right)^{|\beta|^2/2} \ln ^2\!\left(1-\frac{a^2 e^{-2 i \theta }}{|\beta|^2}\right) \, .
\end{align}
Now we approximate $\ln^2(1-x)\approx x^2 + O(x^3)$ for $x\ll 1$, thus
\begin{align}\label{eq:higherorder_approx}
{\rm E}[f(m)]\approx& 	\left(1-\frac{e^{-2i \theta}\hat{a}^2}{|\beta|^2}\right)^{|\beta|^2/2} \left [1 +
\frac{1}{ 4|\beta|^2}  \left(a^2 e^{-2 i \theta }\right)^2 \right ],
\end{align}
which shows the leading order correction is $O(1/|\beta|^2)$ as claimed. Thus the large LO limit must ensure that $\expt{{\rm signal}|a^4|{\rm signal}}\ll 4 |\beta|^2 $, which agrees with the results in Sec. 4.4. of Ref.~\cite{Tyc_2004}; readers interested in a further discussion should consult  Ref.~\cite{Tyc_2004}.
%https://en.wikipedia.org/wiki/Taylor_expansions_for_the_moments_of_functions_of_random_variables

So we can be completely comfortable with the approximation we investigate the variance of $f(X)$ as well. The expansion  is \cite{benaroya2005}
$
{\rm Var}\left[f(X)\right]\approx 
\big(f'({\rm E}\left[X\right])\big)^2 {\rm Var}\left[X\right] 
$.
It turns out that it scales as $1/|\beta|^2$ so in the limit $|\beta|\rightarrow \infty$ the variance becomes zero.

%-------------------------------------------------------------
\section{When \texorpdfstring{$m>n$}{m>n}}\label{app:mgn}
%-------------------------------------------------------------

\subsection{Coherent state local oscillator}\label{app:mgn_cs}
Consider $m > n$  and pull a factor of $(1-a/\beta)^n$ out of \cref{eq:exact_kraus} to arrive at
\begin{align}
    	M_{n,m}^{(\beta)} =&
	\bra{0} 
		\left(1-\frac{\hat{a}}{\beta}\right)^{m-n} 
		\left(1-\frac{\hat{a}^2}{\beta^2}\right)^n 
		\frac{e^{-|\beta|^2/4}}{\sqrt{n!}} 
		\left(\frac{\beta}{\sqrt{2}} \right)^n
		\frac{e^{-|\beta|^2/4}}{\sqrt{m!}} 
		\left(\frac{-\beta}{\sqrt{2}} \right)^m. 
\end{align}
Using the same definition of $x$ from the main text means that
the exponent will have a minus sign which in the $\beta \rightarrow \infty$
limit results in the same expression.  The second operator term has the
exponent changed to $n$ which has the same peak in it's distribution and so
again, results in the same asymptotic limit.

\subsection{Superposition state local oscillator}\label{app:mgn_sup}

Had we chosen that $m > n$ after equation~\ref{eq:exactcat_Kraus}, then this would change equation~\ref{eq:progcat} to become
\begin{align}\label{eq:progcat_reversed}
&M_{n,m}^{[\beta]_\pm} =
\bra{0} 
\frac{ e^{-|\beta|^2/2} } {2^{(n+m)/2} \sqrt{n!m!} \N_\pm(\beta)}
\left(1-\frac{\hat{a}^2}{\beta^2}\right)^n %\nonumber \\
\left[ \beta^n(-\beta)^m 	\left(1+\frac{\hat{a}}{\beta}\right)^{m-n} \pm 
(-\beta)^n \beta^m 	\left(1-\frac{\hat{a}}{\beta}\right)^{m-n}\right].
\end{align}
If now variables are relabelled to swap $n$ and $m$ 
\begin{align}\label{eq:progcat_second}
&M_{n,m}^{[\beta]_\pm} =
\bra{0} 
\frac{ e^{-|\beta|^2/2} } {2^{(n+m)/2} \sqrt{n!m!} \N_\pm(\beta)}
\left(1-\frac{\hat{a}^2}{\beta^2}\right)^m 
\left[ (-1)^n \beta^{m+n}	\left(1+\frac{\hat{a}}{\beta}\right)^{n-m} \pm 
(-1)^m \beta^{m+n}\left(1-\frac{\hat{a}}{\beta}\right)^{n-m}\right].
\end{align}
This equation, up to a global phase factor, only differs from \cref{eq:progcat} in the superposition phase depending on the parity of $n-m$.  Therefore, as the analysis presented in the main text incorporates both superposition phases, it actually also covers the case of $m>n$ if the information as to which superposition phase applies is incorporated.

%-------------------------------------------------------------
\section{Quadrature Eigenstates}\label{app:quadeig}
%-------------------------------------------------------------
This appendix has two parts. In \cref{app:eig} we show that 
\begin{align}
\label{eq:quad_eigenstate}
\ket{x_\varphi} = 
\frac{e^{-x^2/2}}{\pi^{1/4}} 
e^{\sqrt{2}x \chi a^\dagger}
e^{-\chi^2{a^\dagger}^2/2} 
\ket{0},
\end{align}
with  $\chi = e^{-i\theta}$ (for consistency with the LO in the main text in \cref{eq:conv}), is an eigenstate of
\begin{align}
Q(\varphi) =  (e^{-i \varphi} a + e^{i \varphi} a^\dagger )/\sqrt{2}
\end{align}
with eigenvalue $x$, that is
\begin{align}
    Q(\varphi) \ket{x_\varphi} = x \ket{x_\varphi}.
\end{align}
Then in \cref{app:innerprod} we show that 
\begin{align}
 \ip{x'_\varphi}{x_\varphi} =  \delta(x-x^\prime).
\end{align}

%-------------------------------------------------------------
\subsection{Eigenstates}\label{app:eig}
%-------------------------------------------------------------

Our method is inspired by Ref.~\cite{SotoMoya2013}. We start by considering the operator $Q(\varphi)$ acting on $\ket{x_\varphi}$,
\begin{align}
Q(\varphi)\ket{x_\varphi} = 
\frac{e^{-x^2/2}}{\pi^{1/4}}  
\frac{e^{-i \varphi} a + e^{i \varphi} a^\dagger }{\sqrt{2}} 
e^{-\chi^2{a^\dagger}^2/2} 
e^{\sqrt{2}x \chi a^\dagger}
\ket{0}.\nonumber
\end{align}
Note that two operator exponential commute and so their order does not matter.
Left multiply the above equation using a resolution of the the identity 
\begin{align}
I =
e^{-\chi^2{a^\dagger}^2/2} e^{\sqrt{2}x \chi a^\dagger}
e^{-\sqrt{2}x \chi a^\dagger} e^{\chi^2{a^\dagger}^2/2},
\end{align}
which we will then try to remove the quadrature operator by evaluating the conjugations that surround it.  We use of the following commutation relation,  $[a, f(a^\dagger)] = \frac{\partial}{\partial a^\dagger} f(a^\dagger)$, for any smooth function $f$.  If $f(a^\dagger)=e^{-\chi^1 {a^\dagger}^2/2}$, then
\begin{align}
    a e^{-\chi^2 {a^\dagger}^2/2} - e^{-\chi^2 {a^\dagger}^2/2} a = - \chi^2 {a^\dagger}  e^{-\chi^2 {a^\dagger}^2/2} 
\end{align}
Defining an operator $G$ which is a conjugated version of the quadrature operator using only the first part of the identity resolution, and using the above equation gives,
\begin{align}
G = e^{\chi^2{a^\dagger}^2/2} Q(\varphi) e^{-\chi^2{a^\dagger}^2/2} 
&= e^{\chi^2{a^\dagger}^2/2}\left ( \frac{e^{-i \varphi} a + e^{i \varphi} a^\dagger }{\sqrt{2}}  \right )e^{-\chi^2{a^\dagger}^2/2}\\
%&=  \frac{e^{i \varphi}  }{\sqrt{2}} a^\dagger + \frac{e^{-i \varphi}}{\sqrt{2}}e^{\chi^2{a^\dagger}^2/2} a e^{-\chi^2{a^\dagger}^2/2}\\
&=  \frac{e^{i \varphi}  }{\sqrt{2}} a^\dagger + \frac{e^{-i \varphi}}{\sqrt{2}} (a - \chi^2{a^\dagger} )\\
&=  \frac{a^\dagger   }{\sqrt{2}}(e^{i\varphi}-\chi^2e^{-i\varphi}) + \frac{e^{-i \varphi}}{\sqrt{2}} a . 
\end{align}
Next, using the same derivative commutation relation, but with the choice $f(a^\dagger)=e^{\sqrt{2}x \chi {a^\dagger}}$
\begin{align}
    a e^{\sqrt{2}x \chi {a^\dagger}} - e^{\sqrt{2}x \chi {a^\dagger}} a =
    \sqrt{2}x\chi e^{\sqrt{2}x \chi {a^\dagger}},
\end{align}
the $G$ operator can be conjugated again to give the operator $H$,
\begin{align}
H = e^{-\sqrt{2}x\chi a^\dagger} G  e^{\sqrt{2}x\chi a^\dagger}
&=  e^{-\sqrt{2}x\chi a^\dagger} \left( \frac{a^\dagger   }{\sqrt{2}}(e^{i\varphi}-\chi^2e^{-i\varphi}) + \frac{e^{-i \varphi}}{\sqrt{2}} a  \right )e^{\sqrt{2}x\chi a^\dagger}\\
%&=   \frac{a^\dagger   }{\sqrt{2}}(e^{i\varphi}-\chi^2e^{-i\varphi}) + \frac{e^{-i \varphi}}{\sqrt{2}}e^{-\sqrt{2}x\chi a^\dagger} a  e^{\sqrt{2}x\chi a^\dagger}\\
&=   \frac{a^\dagger   }{\sqrt{2}}(e^{i\varphi}-\chi^2e^{-i\varphi}) + \frac{e^{-i \varphi}}{\sqrt{2}}( a +  \sqrt{2}x\chi ).
\end{align}
Now we set $\theta = - \varphi$ or $\chi=e^{i\theta}$ (which gives consistency with \cref{eq:quadeig}) which sets the $a^\dagger$ term to zero, and recall that $a\ket{0} = 0$
\begin{align}\label{eq:x_eigapp}
Q(\varphi)\ket{x_\varphi} =
e^{-\chi^2{a^\dagger}^2/2} e^{\sqrt{2}x \chi a^\dagger} H \ket{0} =
x e^{-\chi^2{a^\dagger}^2/2} e^{\sqrt{2}x \chi a^\dagger} \ket{0} =
x \ket{x_\varphi},
\end{align}
which shows that the state in~\ref{eq:quad_eigenstate} is an eigenstate of the quadrature operator.  All that remains is to normalise this state to give the desired result.

%-------------------------------------------------------------
\subsection{Inner product}\label{app:innerprod}
%-------------------------------------------------------------
In this appendix we compute the normalisation of two squeezed states $\ket{x_\varphi}$ and $\ket{x'_\varphi}$, see \cref{eq:quadeig}. We will show that
\begin{equation}
\ip{x'_\varphi}{x_\varphi} =\frac{e^{-(x'^2+x^2)/2}}{\sqrt{\pi}} 
\bra{0}
e^{-{\chi^*}^2{a}^2/2} 
e^{\sqrt{2}{x^\prime} {\chi^*} a}
e^{\sqrt{2}x \chi a^\dagger}
e^{-\chi^2{a^\dagger}^2/2} 
\ket{0}
=
\delta(x-x^\prime) =\ip{x'}{x}, 
\end{equation}
where $x$ and $x'$ are quadrature eigenstates.
Note that the normalisation of an $x$ eigenstate is very different to any squeezed state.  So we should not expect $\ket{\psi}=S(r)\ket{0}$, with length $\sqrt{|\ip{\psi}{\psi}|^2}=1$, to have any relationship to $\ket{x}$ which is unbounded and behaves like a delta function.

To simplify this expression we will insert the identity operator in the coherent state basis twice, i.e.
\begin{align}
    \frac{1}{\pi} \int d^2 \alpha \ket{\alpha}\bra{\alpha} = I.
\end{align}

If we define $\mathcal N = e^{-x^2/2}/\pi^{1/4} $ and  $\mathcal N' =  e^{-x'^2/2}/ \pi^{1/4} $ and $\mathcal N^2 :=\mathcal N' \mathcal N$.  Then the inner product becomes
\begin{align*}
\ip{x'_\varphi}{x_\varphi} 
&= \mathcal N^2 \bra{0}  e^{-{\chi^*}^2 a^2/2} e^{\sqrt{2} x^\prime \chi^* a}
    e^{\sqrt{2} x \chi a^\dagger} e^{-\chi^2{a^\dagger}^2/2} \ket{0} \\
    &= \frac{\mathcal N^2}{\pi^2} \int d^2 \alpha d^2 \beta\,
        \bra{0} e^{-{\chi^*}^2 a^2/2} \ket{\alpha}\bra{\alpha} 
        e^{\sqrt{2} x^\prime \chi^* a} e^{\sqrt{2} x \chi a^\dagger} 
        \ket{\beta}\bra{\beta} e^{-\chi^2 {a^\dagger}^2/2} \ket{0}. 
\end{align*}
Next we re-order (in normal order) or commute exponentials of $a$ and $a^\dagger$ i.e.
\begin{align}
    e^{\sqrt{2}x^\prime\chi^* a} e^{\sqrt{2}x \chi a^\dagger} =
    e^{2x x^\prime} e^{\sqrt{2}x \chi a^\dagger} e^{\sqrt{2}x^\prime \chi^* a},
\end{align}
where $\chi=e^{-i\theta}$. Doing so and simplifying gives
\begin{align}
\begin{split}
\ip{x'_\varphi}{x_\varphi} 
    &= \frac{\mathcal N^2}{\pi^2} \int d^2 \alpha d^2 \beta\,
        e^{-|\alpha|^2/2} e^{-{\chi^*}^2 \alpha^2/2} 
        e^{-|\beta|^2/2} e^{-\chi^2 {\beta^*}^2/2} 
        e^{2 x x^\prime} 
        \bra{\alpha} e^{\sqrt{2} x \chi a^\dagger} 
        e^{\sqrt{2} x^\prime \chi^* a} \ket{\beta} \\
    &=  \frac{\mathcal N^2}{\pi^2} \int d^2 \alpha d^2 \beta
        e^{-|\alpha|^2/2} e^{-{\chi^*}^2 \alpha^2/2} 
        e^{-|\beta|^2/2} e^{- \chi^2 {\beta^*}^2/2}
        e^{2 x x^\prime} e^{\sqrt{2} x \chi \alpha^*} 
        e^{\sqrt{2} x^\prime \chi^* \beta} 
        e^{-\frac{1}{2}(|\alpha|^2 + |\beta|^2 - 2 \alpha^* \beta)} \\
    &=  e^{2 x x^\prime} \frac{\mathcal N^2}{\pi^2} 
        \int d^2 \alpha 
        e^{-|\alpha|^2} e^{-{\chi^*}^2 \alpha^2/2} 
        e^{\sqrt{2} x \chi \alpha^*} 
        \int d^2 \beta
        e^{-|\beta|^2} e^{-\chi^2 {\beta^*}^2/2}  
        e^{(\sqrt{2} \chi^* x^\prime+\alpha^*) \beta}\\
    &= e^{2 x x^\prime} \frac{\mathcal N^2}{\pi^2} \int d^2 \alpha 
        e^{-|\alpha|^2} e^{-{\chi^*}^2\alpha^2/2} e^{\sqrt{2} x \chi \alpha^*}
        \pi e^{-\frac{\chi^2}{2}(\sqrt{2}{\chi^*}x^\prime + \alpha^*)^2} ,
        \end{split}
\end{align}
where we used the integral $\int d^2\gamma e^{-|\beta|^2}e^{-a \beta^{*2}}e^{c \beta}= \pi e^{-a c^2}$ to arrive at the last line. Further manipulations give
\begin{align}
\begin{split}
\ip{x'_\varphi}{x_\varphi} 
    &= e^{2 x x^\prime} e^{-{x^\prime}^2} \frac{\mathcal N^2 }{\pi} 
        \int d^2 \alpha  e^{-|\alpha|^2} 
        e^{-{\chi^*}^2 \alpha^2/2}  e^{- \chi^2 {\alpha^*}^2/2}
        e^{\sqrt{2} (x-x^\prime) \chi \alpha^*} \\
    &= e^{2 x x^\prime} e^{-{x^\prime}^2} \frac{\mathcal N^2 }{\pi} 
        \int d^2 \alpha  e^{-\frac{1}{2}(\chi^* \alpha + \chi \alpha^*)^2}
        e^{\sqrt{2} (x-x^\prime) \chi \alpha^*} \\
    &= e^{2 x x^\prime} e^{-{x^\prime}^2}  e^{\frac{1}{4}(x-x^\prime)^2}
    \frac{\mathcal N^2 }{\sqrt{2 \pi}} 
    \int d \Im\{\chi^* \alpha\}  e^{-i \sqrt{2} (x-x^\prime) \Im\{\chi^* \alpha\}}  \\
    &= e^{2 x x^\prime} e^{-{x^\prime}^2}  e^{\frac{1}{4}(x-x^\prime)^2}
    \frac{\mathcal N^2 }{2 \sqrt{\pi}} 
    \int d k  e^{-i k (x-x^\prime)}\\
    &= e^{2 x x^\prime} e^{-{x^\prime}^2}  e^{\frac{1}{4}(x-x^\prime)^2} \frac{e^{-x'^2/2} e^{-x^2/2}}{\sqrt{\pi}}
    \sqrt{\pi} \delta(x-x^\prime).
     \end{split}
\end{align}
As the last part is $\delta(x-x^\prime)$, only the value at $x=x^\prime$ matters.  Hence this final expression is equivalent to
\begin{align}
 \ip{x'_\varphi}{x_\varphi} =  \delta(x-x^\prime).
\end{align}
By multiplying the state by the square-root of the inverse of the pre-factor before the delta function, gives the standard normalisation for position eigenstates.

%-------------------------------------------------------------
\section{Cat state homodyne}\label{app:catclick}
%-------------------------------------------------------------
In this section we use coherent states as a proxy for an 
arbitrary signal states with $\langle n_{\rm signal} \rangle \ll \langle n_{\rm LO} \rangle$. This let's us reason about properties of the click distribution that are due, largely, to the LO.

\begin{align}
M_{n,m}^{[\beta]_\pm}\ket{\alpha} =&
\frac{ e^{-(|\beta|^2+|\alpha|^2)/2} } {2^{(n+m)/2} \sqrt{n!m!} \N_\pm(\beta)}
\left( (\alpha+\beta)^n (\alpha-\beta)^m \pm (\alpha-\beta)^n (\alpha+\beta)^m\right)\\
=&
\frac{1} {\N_\pm(\beta)}
\left(  \frac{(\alpha+\beta)^n } {2^{n/2} \sqrt{n!} }e^{-|\alpha +\beta|^2/2}
\frac{(\alpha-\beta)^m  } {2^{m/2} \sqrt{m!}}e^{-|\alpha -\beta|^2/2}
\pm 
\frac{(\alpha-\beta)^n } {2^{n/2} \sqrt{n!} }e^{-|\alpha -\beta|^2/2}
\frac{(\alpha+\beta)^m  } {2^{m/2} \sqrt{m!}}e^{-|\alpha +\beta|^2/2}\right)
\end{align}
which is an amplitude. The detection probabilities are
\begin{align}
\Pr(n,m|\pm,\beta,\alpha)=
\frac{1}{ 2(1\pm e^{-2|\beta|^2})} 
&\left [
 \frac{|\alpha+\beta|^{2n} } {2^{n} n! }e^{-|\alpha +\beta|^2}
\frac{|\alpha-\beta|^{2m}  } {2^{m} m! }e^{-|\alpha -\beta|^2} \right .\nonumber \\
&\pm
 \frac{(\alpha+\beta)^n (\alpha^*-\beta^*)^n } {2^n n!} e^{-|\alpha +\beta|^2}
\frac{(\alpha-\beta)^m (\alpha^*+\beta^*)^m } {2^m m!}e^{-|\alpha -\beta|^2}
\nonumber \\
&\pm
 \frac{(\alpha^*+\beta^*)^n (\alpha-\beta)^n } {2^n n!} e^{-|\alpha +\beta|^2}
\frac{(\alpha^*-\beta^*)^m (\alpha+\beta)^m } {2^m m!}e^{-|\alpha -\beta|^2}
\nonumber \\
&\left.+ \frac{|\alpha-\beta|^{2n}}{2^{n}n!} e^{-(|\alpha-\beta|)^2} 
\frac{|\alpha+\beta|^{2m}}{2^{m}m!} e^{-(|\alpha+\beta|)^2} \right ]
\end{align}
which gives probabilities proportional to separate Poisson distributions.

\subsection{Marginal click distribution for \texorpdfstring{$\alpha=0$}{α=0}}

Below we assume that $\beta$ is real, i.e.  $\beta= |\beta|$
\begin{align}
M_{n,m}^{[\beta]_\pm}\ket{0} =&
\frac{ e^{-(|\beta|^2)/2} } {2^{(n+m)/2} \sqrt{n!m!} \N_\pm(\beta)}
\left( (\beta)^n (-\beta)^m \pm (-\beta)^n (\beta)^m\right)
=
\frac{ e^{-(|\beta|^2)/2} } {2^{(n+m)/2} \sqrt{n!m!} \N_\pm(\beta)}
|\beta|^n|\beta|^m\left(  (-1)^m \pm (-1)^n \right)
\end{align}
Thus
\begin{align}
\Pr(n,m|\pm,\beta,0) 
=&
\frac{ e^{-|\beta|^2} } {2^{(n+m)} n!m! \N^2_\pm(\beta)}
|\beta|^{2n}|\beta|^{2m} 2(1 \pm (-1)^{n+m}) ,
\end{align}
where we used $\left(  (-1)^m \pm (-1)^n \right) = 2(1\pm (-1)^{n+m})$.
The marginal over $n$ is
\begin{align}
\Pr(m|\pm,\beta,0)  
&= \sum_{n=0}^\infty\Pr(n,m|\pm,\beta,0) .
\end{align}

Lets do the $+$ superposition case first
\begin{subequations}
\begin{align}
\Pr(m|+,\beta,0)  
= \sum_{n=0}^\infty\Pr(n,m|+,\beta,0) &=
\frac{ e^{-|\beta|^2/2} } {m!} 
\left(\frac{|\beta|^{2}}{2}\right )^m  \frac{e^{|\beta|^2}+(-1)^m}{e^{-|\beta|^2}+e^{|\beta|^2}}\\
&= \frac{ e^{-|\beta|^2/2} } {m!} 
\left(\frac{|\beta|^{2}}{2}\right )^m  \underbrace{\frac{ (e^{|\beta|^2}+(-1)^m)}{2 \cosh{|\beta|^2}}}_{=1}\\
&= \frac{ e^{-|\beta|^2/2} } {m!} 
\left(\frac{|\beta|^{2}}{2}\right )^m 
\end{align}
\end{subequations}
where we have used $m$ is even for the $+$ superposition. {\em Thus we have a Poisson distribution with mean and variance equal to $|\beta|^2/2$. }

Now we do the minus superposition ``$-$'' case 
\begin{subequations}
\begin{align}
\Pr(m|-,\beta,0)  
= \sum_{n=0}^\infty\Pr(n,m|-,\beta,0) &=
\frac{ e^{-|\beta|^2} } {m!} 
\left(\frac{|\beta|^{2}}{2}\right )^m  \frac{e^{|\beta|^2}+(-1)^{m+1}}{-e^{-|\beta|^2}+e^{|\beta|^2}}\\
&= \frac{ e^{-|\beta|^2/2} } {m!} 
\left(\frac{|\beta|^{2}}{2}\right )^m  \underbrace{\frac{ e^{-|\beta|^2/2}(e^{|\beta|^2}+(-1)^m)}{2 \sinh{|\beta|^2}}}_{=1}\,.
\end{align}
\end{subequations}
The marginals $\Pr(n|\pm,\beta,0)$ look identical with the role of $m$ and $n$ reversed.

% ============================
\section{Small and Moderate local oscillators}\label{apx:LO_size}
% ============================
Here we give some numerical evidence that the local oscillator does not have to be very large to see the effects we describe in the main text. In \cref{fig:lo_size} we consider a fixed coherent state signal, $\alpha =1.6$ and vary the strength of the LO $\beta$. When $\beta$ is less than one the two peaks are not discernible (not shown). As $\beta$ increases (left to right in the figure) the bimodal distribution of the outcomes becomes increasingly apparent. When $\beta = 1.85$ the two modes of the distributions are well separated but seem to be asymmetric about $|x|\approx 2$. For $\beta\ge 2.6$ the asymmetry  seems to be consistent with the LO of $\beta = 5 $ in middle column of \cref{fig:coherent_state_plots}. Again we point the reader to results in Sec. 4.4. of Ref.~\cite{Tyc_2004} for a fuller discussion of the large LO limit.

 \newcommand{\apxfiguresize}{0.31\columnwidth}
\begin{figure}[htb]
    \centering
    \includegraphics[width=\apxfiguresize]{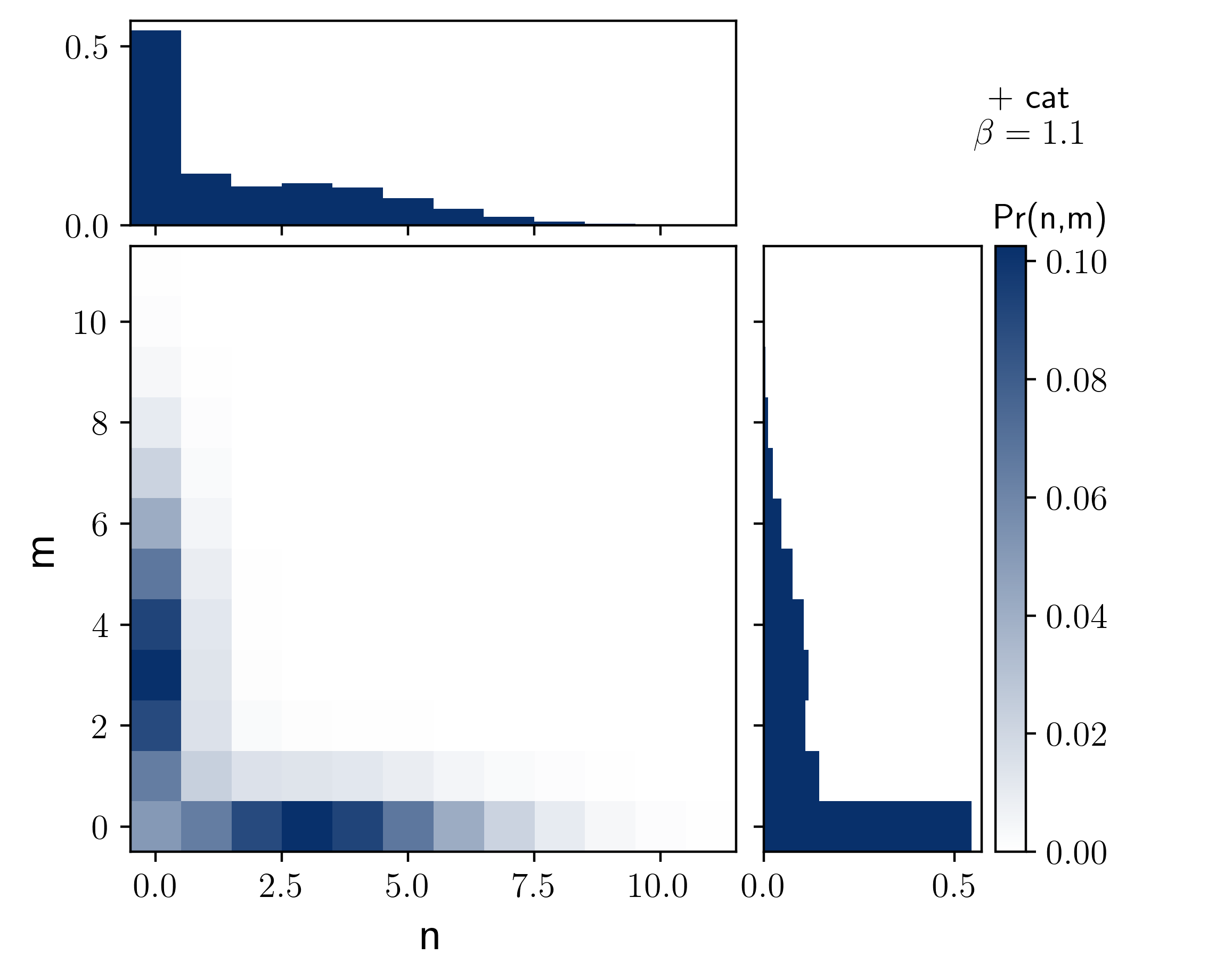}
    \includegraphics[width=\apxfiguresize]{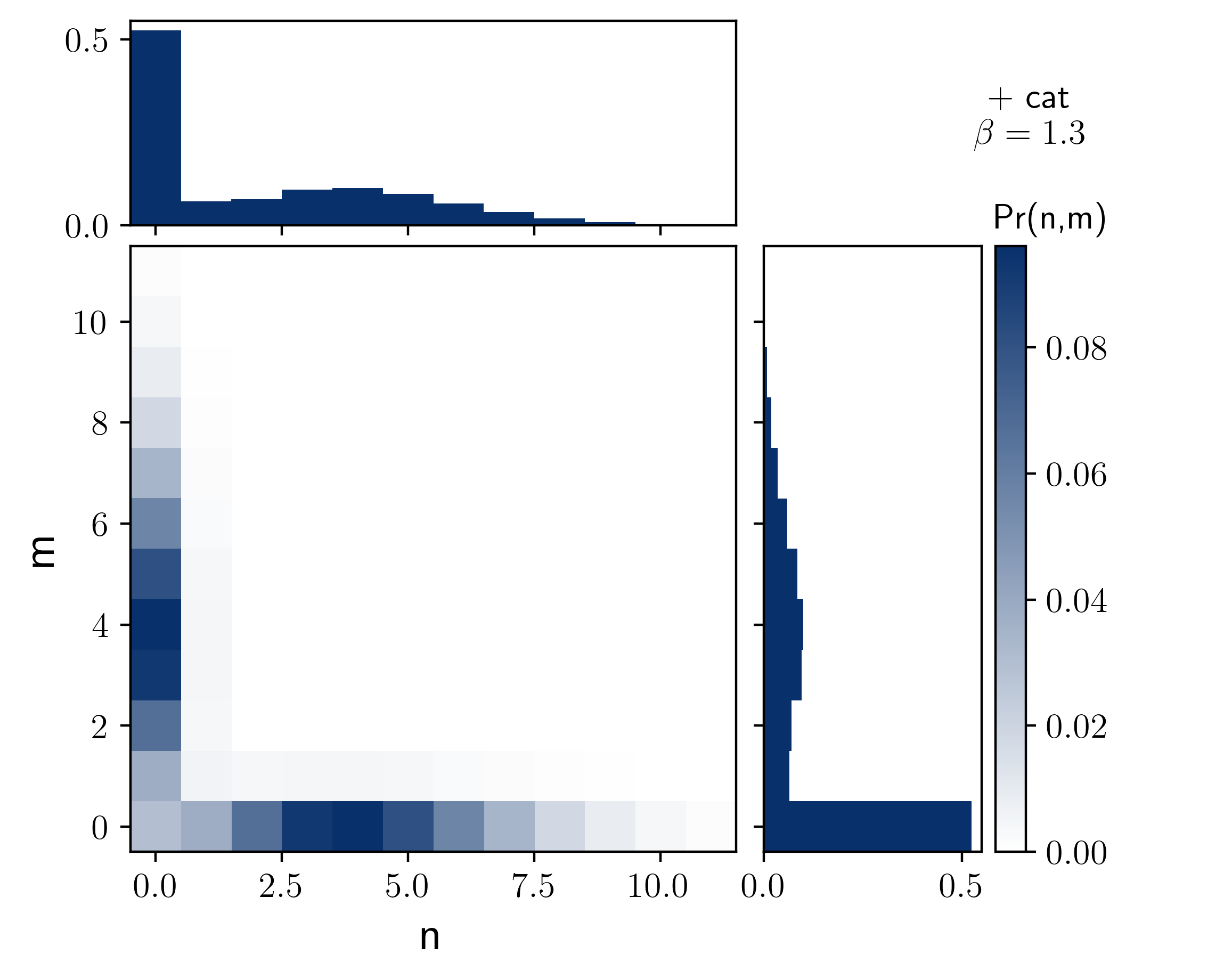}
    \includegraphics[width=\apxfiguresize]{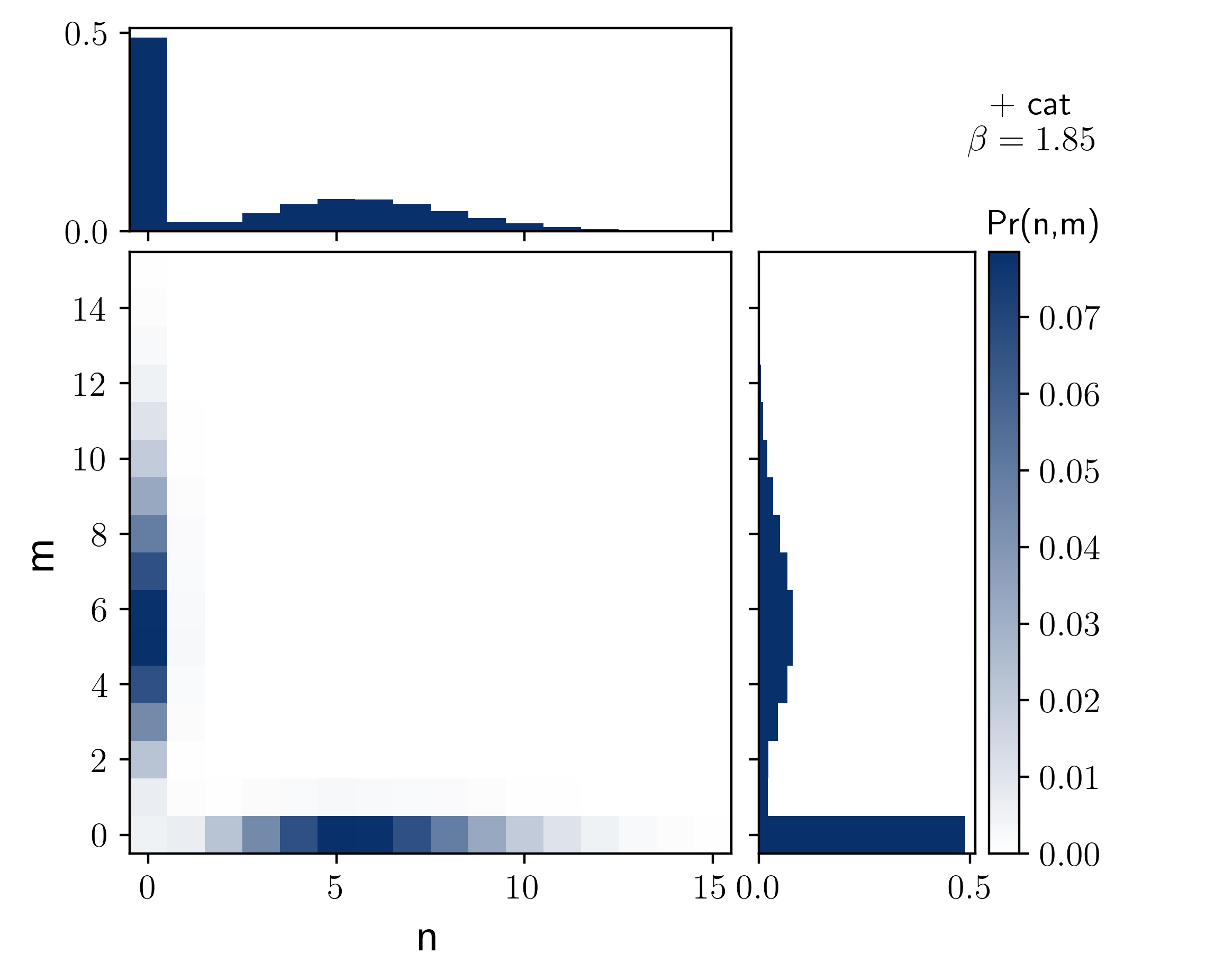}\\
    \includegraphics[width=\apxfiguresize]{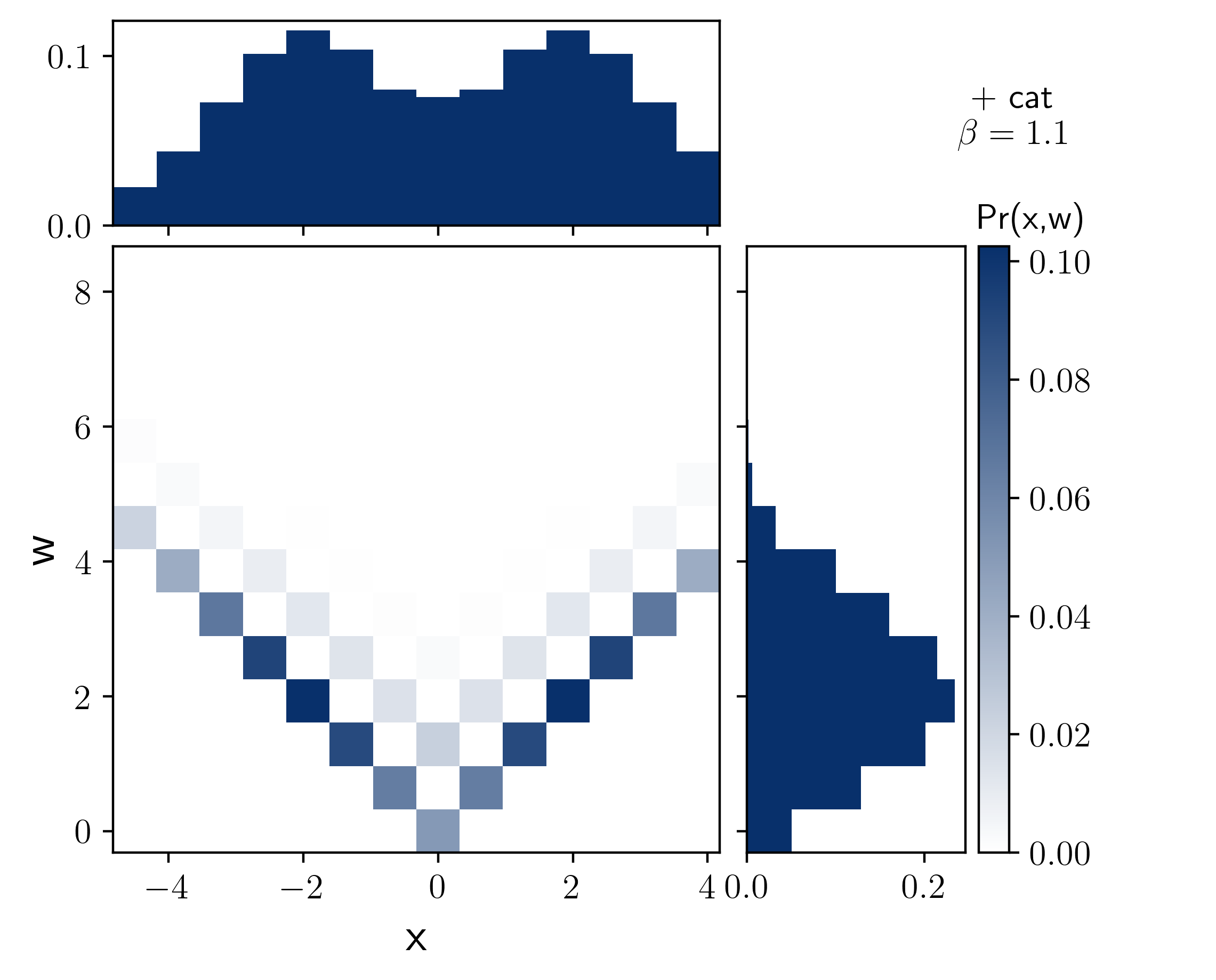}
    \includegraphics[width=\apxfiguresize]{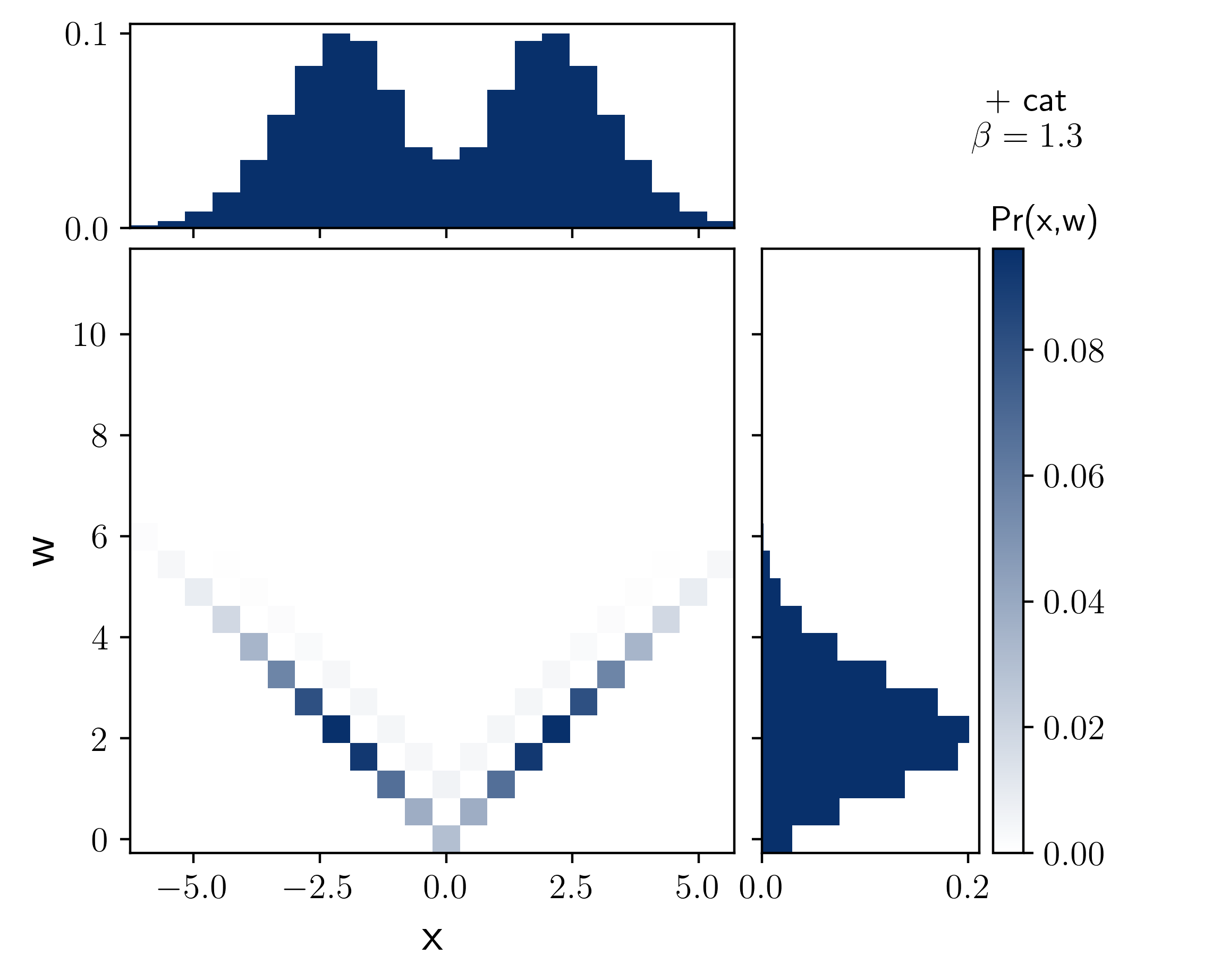}
    \includegraphics[width=\apxfiguresize]{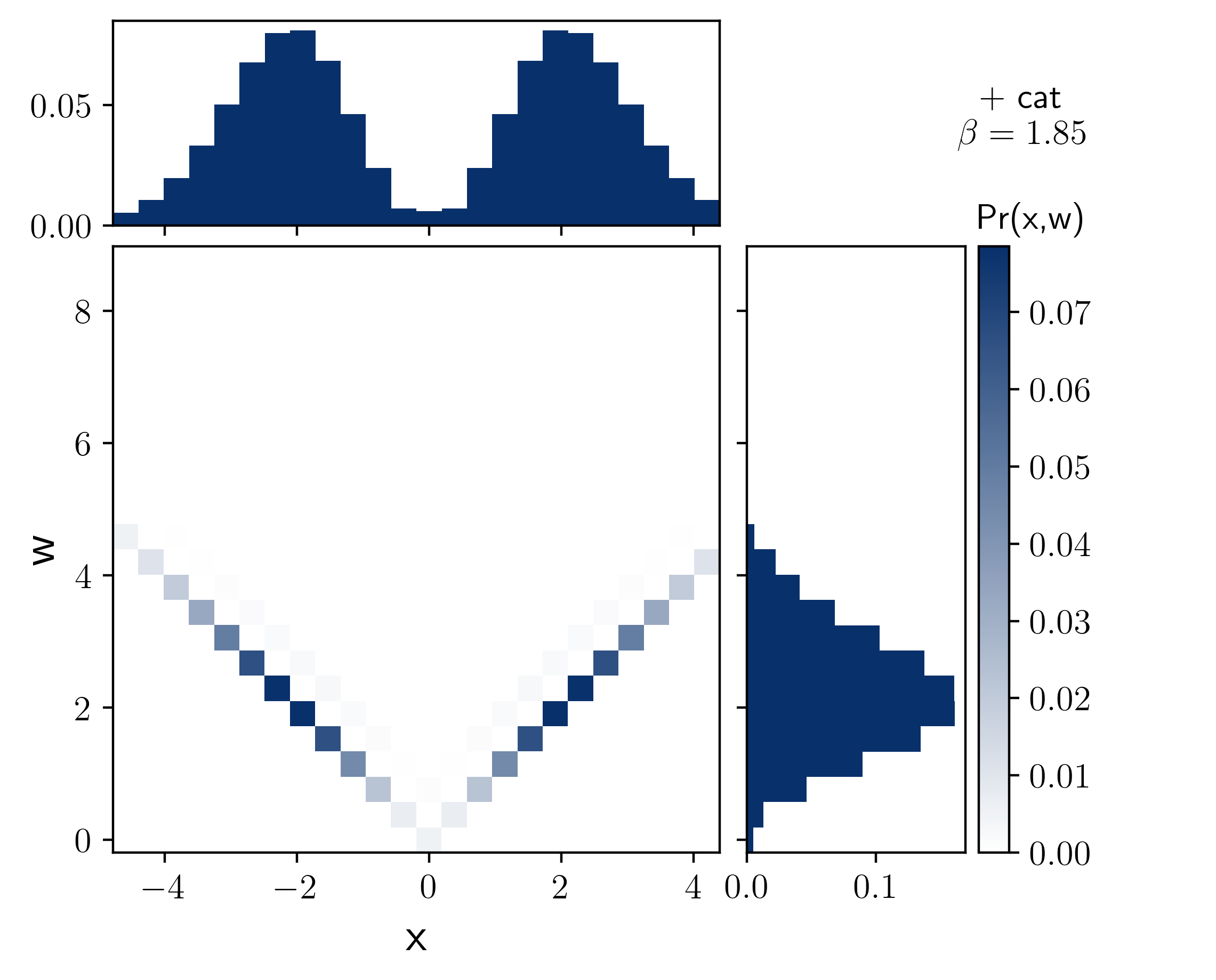}
    \caption{ The effect of the local oscillator size on the click  distributions, in all figures the signal is a $\alpha = 1.6$ coherent state. (column 1) $\beta = 1.1$, (column 2) $\beta=1.3$, and (column 3) $\beta= 1.85$. (Row 1) Original click distributions and marginal distributions for $n$ and $m$. (Row 2) Sum $w=(n+m)/(\sqrt{2}\beta)$ and difference $x=(n-m)/(\sqrt{2}\beta)$ variables and corresponding marginal distributions. }\label{fig:lo_size}
\end{figure}

% =============================
\section{Fock state signal}\label{apx:fock}
% ============================
To further simplify \cref{eq:fockfragment} in the main text, consider the first sub-term
\begin{align}
    \bra{0}(\hat{a}+\beta)^n (\hat{a}-\beta)^m \ket{\focksym}.
\end{align}
The annihilation operator is the only operator within this expression and hence all the operators commute.  Therefore the standard binomial expansion can be used,
\begin{align}
    (a+x)^n = \sum_{k=0}^n \binom{n}{k} a^{n-k} x^k .
\end{align}
This gives
\begin{align}
    (a+\beta)^n (a-\beta)^m \ket{\focksym} = 
    \sum_{k=0}^m \sum_{k^\prime=0}^n
    \binom{m}{k}\binom{n}{k^\prime} (-1)^k \beta^{k+k^\prime}
    \sqrt{\frac{\focksym!}{(\focksym-(m-k)-(n-k^\prime))!}}
    \ket{\focksym-(m-k)-(n-k^\prime)}.
\end{align}
where any negative values within the ket are equivalent to the zero vector. The expression closed with a $\bra{0}$ is then 
\begin{equation}
    \bra{0}(\hat{a}+\beta)^n (\hat{a}-\beta)^m \ket{\focksym} =
    \sum_{k=0}^m 
    \binom{m}{k} \binom{n}{n+m-\focksym -k} (-1)^k \beta^{n+m- \focksym} \sqrt{\focksym !},
\end{equation}
provided $n+m\geq \focksym $ otherwise the expression is zero.  This condition occurs because there is no loss considered in the model and hence if $\focksym$ photons are injected into the detector, at the very least they must all be detected.  Additional photons can arise from the local oscillator which introduces the terms involving the $\beta$.  The series has the form of a ordinary hypergeometric function evaluated at $-1$ and hence can be written
\begin{equation}
    \bra{0}(\hat{a}+\beta)^n (\hat{a}-\beta)^m \ket{\focksym} = 
    \beta^{n+m-\focksym} \sqrt{\focksym!}  
    \begin{cases}
    \binom{n}{m+n-\focksym} \ _2F_1 (-m, -m-n+\focksym; 1-m+\focksym; -1) & m \leq \focksym \\
    \binom{m}{m-\focksym} (-1)^{m-\focksym} \ _2F_1 (-n, -\focksym; 1+m-\focksym; -1) & m > \focksym
    \end{cases}
\end{equation}
again, provided that $n+m \geq \focksym$.  The leading two parameters, given this inequality are always negative which corresponds to the finite sum of terms in the defining series.  The second terms, given the constraints on each branch, are always greater than or equal to one which ensures no singular points.  %This closed form helps mainly with numerical computation but offers little insight into the general functional properties.  

% ============================
\section{Details of finite squeezing}\label{apx:EPR_twomode_squeeze}
% ============================

Let's consider the case where the Kraus operator is $M_{q,+}\propto\bra{q} + \bra{-q}$ (positive parity), where $q$ is the measurement outcome. (The negative parity case where $M_{q,-}\propto\bra{q} - \bra{-q}$ is similar.) Then the unnormalized post measurement state is proportional to 
\begin{equation} 
M_{q,+}\otimes \Id\ket{\psi}_{\rm TMSV} \propto \int dx_b \sqrt{\frac{2}{\pi}}  \left [
e^{-e^{-2r}(q+x_b)^2/2}e^{-e^{2r}(q-x_b)^2/2} +
e^{-e^{-2r}(-q+x_b)^2/2}e^{-e^{2r}(q+x_b)^2/2}
\right ] \ket{x_b},
\end{equation}
here we have used $\ip{\pm q}{x_a} = \delta(x_a\pm q)$ and integrated over $x_a$. We normalize this position wavefunction  and do some algebraic simplification on it and find
\begin{equation} 
\psi(x_b|q,+,r)=\left (\frac{\cosh 2 r}{2 \pi }\right ) ^{1/4} 
\frac{e^{q^2 \cosh 2 r}}{\sqrt{e^{2 q^2 \sinh 2 r \tanh 2 r}+1}}
\left [e^{-  (q^2+x_b^2) \cosh 2 r -2 q x_b \sinh 2 r } + e^{-  (q^2+x_b^2) \cosh 2 r +2 q x_b \sinh 2 r } \right ].
\end{equation}
To further simplify we can complete the square on the exponents, i.e., 
\begin{subequations}
\begin{align}
    -  (q^2+x_b^2) \cosh 2 r -2 q x_b \sinh 2 r  &= -\cosh 2r (x_b + q \tanh 2r)^2 - q^2 \sech 2r \\
      -  (q^2+x_b^2) \cosh 2 r +2 q x_b \sinh 2 r & = -\cosh 2r (x_b - q \tanh 2r)^2 - q^2 \sech 2r
\end{align}
\end{subequations}
which gives the normalized position wavefunction $\psi(x|q,+,r)$ after obtaining measurement outcome ${q,+}$ with squeezing $r$ given in \cref{eq:postmeas_finite_squeeze}.
 
To get an idea about the likelihood of preparing various states as a function of the measurement outcome $q$ we need to compute the probability for obtaining outcome $q$. We consider a simplified version of the full POVM in \cref{eq:catKraus_nointegral} 
\begin{equation}
 dq\, E_{q,\pm} = \frac{dq}{4}\Big ( \op{q}{q}\otimes I_b \pm \op{-q}{q}\otimes I_b \pm \op{q}{-q} \otimes I_b+ \op{-q}{-q} \otimes I_b \Big ) , 
\end{equation}
where $\sum_\pm \int dq\, E_{q,\pm} = \Id \otimes \Id$.
The probability density for obtaining outcome $q$ is
\begin{subequations}
\begin{align}
dq \Pr(q|r) &=\sum_{s \in \{+,- \}} \, _{\rm TMSV}\bra{\psi} E_{q,s}\ket{\psi}_{\rm TMSV}\\
&=\frac{dq}{2} \, _{\rm TMSV}\bra{\psi} \op{q}{q}\otimes I_b +\op{-q}{-q} \otimes I_b  \ket{\psi}_{\rm TMSV}\\
&= dq \,\int dx_a' \int dx_b' \int dx_a \int dx_b \, \psi(x_a',x_b')   \psi(x_a,x_b)\bra{x_a'}\otimes \bra{x_b'}   \Big (\op{q}{q}\otimes I_b \Big) \ket{x_a}\otimes \ket{x_b}.
\end{align} 
\end{subequations}
Performing the integrals and simplifying we find
$ dq\,\Pr(q|r) = dq (2 \sech 2 r /\pi)^{1/2}\exp[-2 q^2 \sech 2 r]$.

\end{widetext}

%%%%%%    The BIB   %%%%%%%
\bibliography{homodyne}
%%%%%%%%%%%%%%%%%%%%%%%%%%%%

\end{document}